\documentclass[manuscript]{aastex}

\slugcomment{Not to appear in Nonlearned J., 45.}

\shorttitle{Ionisation  in atmospheres of very low-mass objects}
\shortauthors{Helling et al.}

\begin{document}


\title{Ionisation in atmospheres of Brown Dwarfs and extrasolar planets
       III. Breakdown conditions for mineral clouds }


\author{Ch. Helling\altaffilmark{1}, M. Jardine\altaffilmark{1}, C. Stark\altaffilmark{1}, D. Diver\altaffilmark{2}, }
\affil{ 1: SUPA, School of Physics \& Astronomy, University of St Andrews, St Andrews,  KY16 9SS, UK\\
2: SUPA, School of Physics and Astronomy, University of Glasgow, Glasgow G12 8QQ, UK}
\email{ch@leap2010.eu}




\begin{abstract}
 Electric discharges were detected directly in the cloudy atmospheres
 of Earth, Jupiter and Saturn, are debatable for Venus, and indirectly
 inferred for Neptune and Uranus in our solar system. Sprites (and
 other types of transient luminous events) have been detected only on
 Earth, and are theoretically predicted for Jupiter, Saturn and Venus.
 Cloud formation is a common phenomenon in ultra-cool atmospheres such
 as in Brown Dwarf and extrasolar planetary atmospheres. Cloud
 particles can be expected to carry considerable charges which may
 trigger discharge events via small-scale processes between individual
 cloud particles (intra-cloud discharges) or large-scale processes
 between clouds (inter-cloud discharges).  We investigate
 electrostatic breakdown characteristics, like critical field
 strengths and critical charge densities per surface, to demonstrate
 under which conditions mineral clouds undergo electric discharge
 events which may trigger or be responsible for sporadic X-ray
 emission. We apply results from our kinetic dust cloud formation
 model that is part of the {\sc Drift-Phoenix} model atmosphere
 simulations. We present a first investigation of the dependence of
 the breakdown conditions in Brown Dwarf and giant gas exoplanets on
 the local gas-phase chemistry, the effective temperature and
 primordial gas-phase metallicity. Our results suggest that different
 intra-cloud discharge processes dominate at different heights inside
 mineral clouds: local coronal (point discharges) and small-scale
 sparks at the bottom region of the cloud where the gas density is
 high, and flow discharges and large-scale sparks near, and maybe
 above, the cloud top. The comparison of the thermal degree of
 ionisation and the number density of cloud particles allows us to
 suggest the efficiency with which discharges will occur in planetary
 atmospheres.
\end{abstract}




\today

\section{Introduction}
Electric discharges, including transient luminous events\footnote{In
  the community of terrestrial atmospheric physics, the term {\it
    transient luminous event} is used to describe sprites, elves, and
  blue-jets only, but not lightning.} or transient discharges in
gases, are non-equilibrium plasmas produced by short-lived electrical
conditions of which lightning and associated sprites are the most
widely known. Lightning and sprites are commonly observed on
Earth. Lightning is also detected or indirectly inferred on most of
the cloud-carrying solar system planets. Observations on Earth show
that cloud particles or droplets trigger such transient discharges in
atmospheric environments (e.g. Yuan et al. 2011).  It is not obvious
that the underlying physical mechanisms for triggering lightning and
sprites are different: they each involve an unbalanced evolution of
accelerated free electron populations, driven by electric potential
differences. How the latter are created from the organisation of
ambient gas/particles may be different, but the underlying science is
not.

Most of the solar system planets form clouds made of liquids  and
  ices, but dependent on local conditions,  mineral hazes
may also form (e.g. Saunders \& Plane 2011). Recent observations of HD
189733b (Pont et al. 2008; Sing et al. 2009, 2011; Gibson et al. 2012)
suggest that extraterrestrial giant gas planets contain small mineral
dust particles high up in their atmospheres. Further examples include
GJ1214b and Kepler 7b.

Prompted by the abundance of discharge processes in clouds on Earth
and in the clouds of solar system planets, we take a first look at the
conditions for such discharge events in extrasolar, planetary and
Brown Dwarf atmospheres. The essential requirements are the presence of
both, an electric field (electric potential difference) and free
electrons. This electric field accelerates the free electrons through
the ambient gas. If they collide with other gas particles sufficiently
frequently and with enough energy to liberate increasingly more
electrons, an electron avalanche is produced. The critical electric
field required for this depends, to some extent, on the composition of
the gas. This avalanche can develop into a self-propagating ionisation
front (streamer) if the self-field of the electrons in the avalanche
is similar to the strength of the applied external field.  If the
avalanche has reached a critical size, it can propagate into areas
with a lower electric field strength in which an avalanche would not
start normally. Two electrical breakdown mechanisms are known to occur
in dielectrics: the conventional breakdown occurring for example in
glow discharges; and the runaway breakdown which is suggested to play
a role in lightning discharges on Earth. The threshold electric field
needed to initiate the avalanche of a runaway breakdown appears one
order of magnitude below that for a conventional breakdown (for more
details see Roussel-Dupre et al. 2008; Marshall et al. 1995).  Each of
the resulting streamers is a highly conducting plasma channel in which
the density of non-thermal electrons is greatly increased, and from
which electrons leak into the ambient gas.

Helling, Jardine \& Mokler (2011) have argued that dust clouds that
are made of mineral particles can be charged e.g. by gas turbulence
induced dust-dust collisions. The cloud particles remain charged for
long enough that discharge events can occur and streamers can be
established. Other possible processes for charge separation, which we
have not considered, are e.g.  different contact potentials of the
material making up the grain, and fractoemission or collisional
charging of polarised grains in an external electric field (P\"ahtz,
Herrman \& Shinbrot 2010). The Coulomb recombination time of cloud
particles by thermal electrons is long enough that electric fields can
establish between passing grains which then can initiate subsequent
electron avalanche processes leading to streamers (Fig. 4 in Helling
et al. 2011). The result is that during the electron avalanche phase
there is an exponential increase in free charges for each
free electron that exists in the gas phase. Numerical experiments that
simulate an avalanche from the initial electron up to and beyond the
avalanche-to-streamer transition demonstrate consistently that a
single electron seeding a streamer can cause the occurrence of
$10^{12}-10^{13}$ secondary non-thermal electrons for a certain time
interval (Dowds et al. 2003, Ebert et al. 2010).

 Laboratory studies suggest that during grain-grain impact,
smaller grains are more negatively charged compared to larger grains
of the same composition (Lacks \& Levandovsky 2007, Forward et
al. 2009, Merrison et al. 2012). A similar conclusion was reached from
volcano ash experiments (Hatakeyama \& Uchikawa 1952, Kikuchi \& Endoh
1982, James et al. 2000). Experimental results for different minerals
 are not available at present. Experiments on wind-driven
entrainment of Martian dust (mainly iron oxides) and Basalt dust
(Na/K/Ca-silicates) by Merrison et al. (2012) suggest average electric
charge values of $10^3\,\ldots\,10^5$ e$^-$/grain.  The exact
  values might, however, be influenced by the experimental set-up
  (Aplin et al. 2012). These experiments were carried out under
pressures between 10\,mbar and 1\,bar. Only a very weak pressure
dependence on the number of charges per grain was seen. No strong
dependence on the grain material was found either, but one order of
magnitude variation is suggested due to grain sizes, morphology and
abundance of grains. They found no evidence for a dependence of dust
electrification on atmospheric composition.

The extrasolar atmospheric environments we are looking at in Brown
Dwarfs and giant gas planets have local temperatures between T$_{\rm
  gas}$= 500K -- 4000K and gas pressures between $p_{\rm gas}\approx
10^{-10}$bar -- 10 bar.  This pressure range comprises the pressure
interval for Mars-dust analog experiments by Merrison et
al. (2012). The resulting degree of ionisation by thermal processes in
Brown Dwarfs and giant gas planet atmospheres is $\chi_{\rm e}=p_{\rm
  e}/p_{\rm gas} \approx 10^{-15}$ -- $10^{-7}$ throughout the atmosphere,
which is very low (Helling et al. 2012), but enough to start an
electron avalanche. The formation of dust by seed formation and bulk
growth takes place in a temperature window of $\sim$500K--$\sim$2100K
and leads to the formation of mineral clouds.  Gravitational
  settling, convective mixing and element depletion are major
 associated processes. Helling, Woitke \& Thi (2008) have shown that
the upper cloud, (i.e. low temperature and low-pressure) will be
dominated by small, dirty (i.e. inclusions of other materials)
silicate grains with inclusions of iron and metal oxides, and the
warmer, denser cloud base by bigger, dirty iron grains with metal
inclusions. The actual size of the cloud particles deviate from
  this mean value according to a height dependent size distribution
  (Fig. 8 in Helling, Woitke \& Thi 2008). The chemical 
composition of the grains as well as their size distribution change with
height inside a cloud in a quasi-stationary environment.

The different scale regimes involved in the formation of atmospheric
clouds suggest the consideration of two different scale regimes with
respect to charge and discharge processes. The large scale regime is
where the cloud particles settle gravitationally (rain out) according
to their mass and size, or where they are transported by
hydrodynamical processes like convection. This will leave the smallest
and potentially more negatively charged particles suspended in the
higher layers, while the larger and more positively charged particles
populate the lower part of the cloud.  The small scale regime
describes the scales at which cloud particles interact during
grain-grain collisions, where colliding particles (droplet) or
individual cloud particles produce point discharges (also called {\it
  corona}) that may start large-scale lightning. This analogy is drawn
from the discharge experiments for Earth conditions where colliding
drops in storms and ice particles can produce point discharges which
are thought to be able to initiate lightning (for more details see
MacGorman \& Rust 1998). A similar scenario is suggested in Merrison
et al. (2012), and was utilised by Farrell et al. (1999) and Michael
et al. (2008) to model Martian dust storms.

What are the conditions to trigger a discharge event which
subsequently increases the local density of free electrons by
non-thermal processes leading potentially to the occurrence of
lightning or associated transient luminous events like sprites?  The
local electric field, $E$, must exceed the ionisation threshold of the
ambient gas, hence a critical field strength, $E_{\rm crit}$, and seed
electrons for the development of an electron avalanche need to be
present. We note that measurements of the actual threshold electric
field strength for a terrestrial lightning flash in a cloud are not
easy to obtain (Marshall et al. 2005).  Marshall et al. (1995)
performed balloon experiments from which they derived electric
breakdown thresholds that led them to hypothesised that lightning
flashes might be initiated by runaway breakdown. This paper therefore
presents a first investigation of the discharge behaviour in
extrasolar atmospheres by studying the following questions.\\
\noindent
-- What electric field strength is needed to allow breakdown
phenomena known from Earth and solar system planets to occur in
extrasolar, planetary objects and Brown Dwarfs?\\
 -- What can we learn
about the breakdown conditions in extrasolar objects, for example breakdown
distances, minimum voltages, or chemistry dependence?\\ 
-- What are the critical conditions for breakdown  such as cloud size, grain size, number of charges per surface? Under which conditions can the critical field strength be produced by grains alone? \\ 
-- How
many thermal electrons are available in a
mineral cloud of a Brown Dwarf or a giant gas planet in order to trigger a streamer event?
How efficiently can discharge events occur and, hence influence the gas phase composition?

To answer these questions, we first outline the basic ideas and
relations for an electrostatic breakdown in gases of different
chemical compositions in Sect.~\ref{s:bdc}. These relations are
evaluated in Sect.~\ref{s:res} for previously calculated model
atmosphere structures of Brown Dwarfs and giant gas planets. We are
using {\sc Drift-Phoenix} model atmosphere structures which are the
result of the solution of the coupled equations of radiative transfer,
convective energy transport (modelled by mixing length theory),
chemical equilibrium (modelled by laws of mass action), hydrostatic
equilibrium, and dust cloud formation (Dehn 2007, Helling et
al. 2008a,b; Witte, Helling \& Hauschildt 2009; Witte et
al. 2011). The dust cloud formation model includes a model for seed
formation (nucleation), surface growth and evaporation of mixed
materials. The effects of gravitational settling (drift) and
convective overshooting on the cloud formation are explicitly included
in our dust model equations (Woitke \& Helling 2003, 2004; Helling \&
Woitke 2006; and Helling, Woitke \& Thi 2008). The results of the {\sc
  Drift-Phoenix} model atmosphere simulations include for example the
gas temperature - pressure structure ($T_{\rm gas}$, $p_{\rm gas}$),
the local gas-phase composition, the local electron number density
($n_{\rm e}$), the number of dust grains ($n_{\rm d}$) and their
height dependent sizes ($a$). These models are determined by the
effective temperature, $T_{\rm eff}$ [K], the surface gravity, g
[cm/s$^2$], and the initial element abundances which are set to the
solar values unless specified otherwise.  The final element abundances
are determined by element depletion due to dust formation. We also
discuss how many charges per surface would be needed to initiate a
field breakdown based on this model output (Sect.~\ref{ss:hmc}), and
we estimate the enrichment rate of the gas-phase by streamer electrons
(Sect.~\ref{ss:disc}). Section~\ref{sec:concl} contains our
conclusions.

\section{Breakdown characteristics}\label{s:bdc}

Electric breakdown is the process of transforming a non-conducting
 substance into a conductor as the result of applying a sufficiently
large electric field. This ionisation state of the bulk gas phase builds up over times
between $10^{-8}\,\ldots\, 10^{-4}$s and is  large enough that it is
accompanied by a light flash. A gas-phase breakdown is a threshold
process, hence, it occurs when the local electric field exceeds a value
characteristic of the special local conditions.

A streamer develops in the wake of a positively charged tail of a
primary avalanche during which an accelerated electron travels through
an ambient gas. If the energy of the accelerated electron exceeds the
ionisation potential of the gas, the growth of a thin ionisation
channel (streamer) between two electrodes results. These electrodes
can be capacitor plates, charge distributions inside clouds, or
grains. The transition between an avalanche process and the streamer
mechanisms is not well documented so far. Raether (1964) suggests that
the streamer mechanism dominates at an electrode distance of 5-6
cm, the exact value of which might change depending on local
  conditions.

The onset of electric field breakdowns has been studied for charge
distributions on capacitor plates for a variety of gases and
characteristic quantities, and the breakdown voltage has been
parametrised for easy use based on lab experiments
(Eq.~\ref{eq:paschenV}-~\ref{eq:paschenE}; e.g. Raizer 1991).
Different electric field strengths inside a capacitor were tested for
different gases.  For practical reasons, the parametrisation of the
breakdown voltage has only been done for terrestrial or solar system
gas mixtures or for homogeneous gases, which may not resemble the gas
composition in an extrasolar planetary or Brown Dwarf atmosphere.

 The resulting Paschen curve, that relates the breakdown voltage with
the product $pd$ (Eqs.~\ref{fig:paschen}, Fig.~\ref{fig:paschen}),
shows that the breakdown voltage of a capacitor, $V_{\rm t}$ [V],
depends only on the material of the cathodes (characterised by Townsend's second
ionisation coefficient $\gamma$), the gas between the capacitor plates
(characterised by the coefficients $A$ [cm$^{-1}$ torr$^{-1}$] and $B$ [V\,cm$^{-1}$
torr$^{-1}$]), and the product between the gas pressure and the capacitor plate
separation, $pd$ [cm torr],
\begin{eqnarray}
\label{eq:paschenV}
V_{\rm t}&=&\frac{B\cdot pd}{C + \ln(pd)} \quad \mbox{with} \quad C=\ln\frac{A}{\ln(\frac{1}{\gamma} + 1)}\\
\label{eq:paschenE}
\frac{E_{\rm t}}{p} &=& \frac{B}{C +  \ln(pd)}
\end{eqnarray}
Townsend's second ionisation coefficient $\gamma$ is the effective
secondary emission coefficient of the cathode and can be calculated as
the ratio between the number of electrons ejected per incident ion.  We
will refrain from doing so and apply numbers given in the literature.
 $\gamma$ varies between $10^{-9}$ for organic cathode materials and
$10^{-2}$ for nickel and copper cathodes (Table 4.10 in Raizer
1991). If $\gamma=1$ there is  one ionisation event per hit. However, the
breakdown voltage becomes independent of the cathode  material
when a spark breakdown or a streamer starts to develop. This is
evident from the fact that lightning in Earth is not triggered by
electron emission of the cloud 'cathodes', the negatively charged part of a cloud.

Qualitatively, the Paschen curve predicts that the breakdown voltage
decreases as the electrode gap decreases because the corresponding
electric field increases. The Paschen curve proved to be accurate for
large gaps and low pressures but it is observed to fail at extremely
low and high $pd$-values. We note that experiments show that the
Paschen curve is not necessarily accurate in describing breakdowns
between electrode spaces of less than 15$\mu$m. Go \& Pohlman (2010)
have therefore derived a model for a modified Paschen curve for
breakdown in micro-scale gaps. In their model, the breakdown voltage
does not have a singularity ($+\infty$) for small gap sizes but instead
decreases further (see their Fig.1). The
observation that $pd=$ const does not hold for small gaps, and $p$ and
$d$ become independent variables. For the small-scale cases
considered, i.e. the distance between two charge centres in the form of
grains, the gap size is just large enough to allow us to apply the
classical Paschen curve formalism (Fig.~\ref{fig:capdist}).

At some gap size, the classical Paschen curve has a minimum, the
Stoletow point, which is calculated from $dV_{\rm t}/d(pd)=0$,
resulting in
\begin{equation}
\label{eq:paschenmin_pd}
pd = \frac{e}{A} \ln(\frac{1}{\gamma} + 1).
\end{equation}
$pd$ is a  local constant  for a given gas composition (expressed by the
constants $A$, $B$) and for a given capacitor surface material
(expressed by the Townsend's second ionisation coefficient $\gamma$)
as long as the gap $d$ is large enough.  The product $pd$ will
therefore be the same for any planet or Brown Dwarf given they have
the same gas composition.  The capacitor plate distance,
$d$, which in our small-scale case is the distance between passing grains, changes
with the local gas pressure, which in our case is a proxy for the vertical position  in the 
atmosphere (Fig.~\ref{fig:capdist}).

Figure~\ref{fig:paschen} shows $V_{\rm t}(pd)$ with the Stoletow point
of these gas-chemistry dependent Paschen curves indicated by octaeder
symbols. Brown dwarfs would be best represented by the Paschen curve
for H$_2$ (brown dashed line) and giant gas planets will very likely
exhibit a more complex gas-phase composition because they form from a
protoplanetary disk at different distances from the host stars
therein. The minimum breakdown voltage $V_{\rm t, min}$ [V] follows by
inserting Eq.~\ref{eq:paschenmin_pd} into Eq.~\ref{eq:paschenV}
(left),
\begin{equation}
\label{eq:paschenmin_Vtmin}
V_{\rm t, min}(pd) = \frac{e^1 B}{A} \ln(\frac{1}{\gamma} + 1).
\end{equation}
Thus, for each gas composition, there is a minimum voltage required to
achieve electric breakdown in the gas. Only then can a streamer or
subsequent lightning develop. It may however be more useful to express
this condition in terms of the electric field strength $\vec{E}$.
With $E=|\vec{E}|=|-\partial \phi(r)/\partial r|$, and $\phi(r)$ the
electric potential and voltage $V=\Delta \phi(r)$ being the potential
difference ($\Delta$) between two points (hence, the units are the
same), it follows $E=V/r$ for a spherical charge distribution at a
radial distance $r$ from the charge $Q$ ($E=V/d$ for a capacitor).
Inserting Eq.~\ref{eq:paschenmin_pd} into Eq.~\ref{eq:paschenE}, the
breakdown or ignition field strength follows with $E=|-\partial
\phi/\partial r|$ from the following relation
\begin{equation}
\label{eq:paschenmin_Ecrmin}
\frac{E_{\rm t, min}}{p} = B \quad\Rightarrow\quad E_{\rm t, min} = B p.
\end{equation}
Hence, $(E_{\rm t, min}/p)$=const is a universal constant of a gas in
the capacitor or the gas exposed to the electric field of some charge
distribution. $E_{\rm t, min}$ is the minimum electric field strength
needed to achieve breakdown. Raizer (1991) and Sentman (2004) have
published $(E_{\rm t, min}/p)$-values for various gases and gas
mixtures potentially representative of the solar system planetary
atmospheres.  These values are based on either experiments or
calculations that take into account details of the interaction
processes of the electron with the ambient gas. For example, Sentman
(2004) deals with gas mixtures by Boltzman modelling of the electron
distribution function for a given gas composition (added to
Table~\ref{tab:AB}).  Hence, details on the acceleration process,
scattering and energy losses/gains are 'hidden' in the measured
material constant. A detailed review of a kinetic approach can be
found in Roussel-Dupre et al. (2008). We summarise the values in
Table~\ref{tab:AB} as some of the original literature is hard to come
by. For more details on composition dependencies consult Chapter 6 in
Meek \& Craggs (1978).

In summary, once the gas composition is defined for a given Brown
Dwarf or gas giant planet, the pressure variation with height defines
the variation of the minimum electric field needed for breakdown in
the atmosphere which we will investigate in more detail below.

\begin{table}[htdp]
\caption{Parametrisation of electric breakdown field according to Paschen curves: Material constants A \& B taken from Raizer (1991) and from Sentman (2004), $pd$ and $V_{\rm t, min}$ calculated from Eqs.~\ref{eq:paschenmin_pd},~\ref{eq:paschenmin_Vtmin}. $\gamma=0.001$}
\begin{center}
\begin{tabular}{r|r|r|| r|r}
               & A & B & $pd_{\rm min}$ & $V_{\rm t, min}$\\
               & [1/(cm\,Torr)]	 & [V/(cm\,Torr)] & [cm\,Torr] & [V]\\
               & (Raizer 1991)  & (Raizer 1991)   & (this paper)  & (this paper)\\ 
 \hline
 He       & 3   &   34 & 4.181 & 142.2\\
 H$_2$ & 5   & 130 & 2.509 & 326.1\\
 N$_2$ & 12 & 342 & 1.045 & 357.5\\
 Air        & 15 & 365&  0.8363 & 305.2\\
 CO$_2$ & 20 & 466 & 0.6272 & 292.3\\
 H$_2$O & 13 & 290 & 0.9649 & 279.8\\
\hline 
              & A & E/p=B & $pd_{\rm min}$& $V_{\rm t, min}$\\
Sentman (2004)& [1/(cm\,Torr)]	 & [V/(cm\,Torr)] & [cm\,Torr] & [V]\\
\hline
Venus {\small (CO$_2$/N$_2$=96.5/3.5)}        & 7.27             & 180            & 2.58 & 465\\
Earth {\small (N$_2$/O$_2$/Ar=78/21/1)}       & 7.44             & 243            & 2.53 & 617\\
Mars  {\small (CO$_2$/N$_2$/Ar=95.5/2.8/1.7)} & 7.23             & 178            & 2.60 & 462\\
Jupiter {\small (H$_2$/He/CH$_4$=89/10.9/0.1)}  & 6.19             & 143            & 3.06 & 434\\
Saturn  {\small (H$_2$/He/CH$_4$=96.3/3.6/0.1)} & 7.46             & 156            & 2.52 & 392\\
Titan \& Tritan  {\small (N$_2$/CH$_4$=95/5)}  & 8.80           & 274            & 2.13 & 585\\
Uranus  {\small (H$_2$/He/CH$_4$=82.5/15.2/2.3)} & 6.47             & 138            & 2.90 & 401\\
Neptune {\small (H$_2$/He/CH$_4$=80/18.5/1.5)}   & 622              & 135            & 3.02 & 408 
\end{tabular}
\end{center}
\label{tab:AB}
\end{table}%

\section{Results}\label{s:res}

The breakdown quantities introduced in the previous section will now
be evaluated for a set of atmosphere structures that were previously
calculated using the {\sc Drift-Phoenix} model atmosphere code by
Witte, Helling \& Hauschildt (2009). These model atmospheres provide
the required input quantities for the following calculations which are
the local gas pressure, $p_{\rm gas}$, and grain sizes, $a$, for each
height in the atmosphere. Two models are studied in detail: $T_{\rm
  eff}=1600$K, log(g)=5.0 as the example for a Brown Dwarf atmosphere,
and $T_{\rm eff}=1600$K, log(g)=3.0 as the example for a
non-irradiated giant gas planet. Both atmospheres have an initially
solar metallicity.  Later in the paper, we compare these results
to models of different effective temperature $T_{\rm eff}=1600, 1900,
2100, 2500$K and  2800K (Fig.~\ref{fig:Qcrit_Teff}), and the initial
metallicity is varied between [M/H]$=-5\,\dots\,0$ with [M/H]=0
indicating the solar values (Fig.~\ref{fig:Qcrit_metal}). The
comparison amongst these models allows us to identify trends
in the electric breakdown quantities in substellar and low-metallicity
atmospheres.

\subsection{Classical breakdown quantities}\label{ssBDQs}
\paragraph{Breakdown distance $d$:} Assuming that a charge separation has built up in a Brown Dwarf or
planetary atmosphere, how far can two imaginary capacitor plates be
apart that a breakdown of the atmospheric gas will occur for an
established electric field?

This is the distance of two capacitor plates, or two otherwise
charge-carrying surfaces, that can build up an electric field between
them, as described by the Paschen minimum where $p_{\rm gas}\cdot d=
2.509$ torr cm (=5007 dyn/cm$^2$) for an H$_2$ gas
(Eq.~\ref{eq:paschenmin_pd}, Fig.~\ref{fig:paschen}). This is the
distance where the electrical breakdown would be easiest according to
Eq.~\ref{eq:paschenmin_Vtmin}. Hence, for $p_{\rm gas}\cdot d$ below
the Paschen minimum, the breakdown voltage increases sharply in the
classical case. However, Go \& Pohlman (2010) show that lower
breakdown voltages should be expected at smaller distances, justifying
our use of the classical Paschen curve as a first step to study
breakdown characteristics. $p_{\rm gas}\cdot d$ is a constant for a
given $A$ [cm$^{-1}$ torr$^{-1}$] and $\gamma$, hence, the same for
all atmospheres of the same chemical composition with respect to the
most abundant gas-phase species. This allows us to calculate the
breakdown distance $d$ for a known $p_{\rm gas}$-structure of a model
atmosphere.

Figure~\ref{fig:capdist} shows the breakdown distance as function of
the local gas pressure calculated for the pressure structure of a
Brown Dwarf (T$_{\rm eff}=1600$K, log(g)=5.0, solar element
abundances, brown line) and a giant gas planet (T$_{\rm eff}=1600$K,
log(g)=3.0, solar element abundances, red lines) assuming that
molecular hydrogen (A[H$_2$]=5 cm$^{-1}$Torr$^{-1}$, Raizer 1991) is
the dominating gas species to be ionised by electrons accelerated in
the electric field. This assumption is appropriate as H$_2$ is the
most abundant molecule in the atmosphere models applied here. The
assumption breaks down if H$_2$ ionises in the hot and dense inner
layers of the atmosphere where clouds can not be thermally stable, and
where atomic hydrogen is the dominating gas component. The Townsend's
2nd ionisation coefficient $\gamma=0.001$ is used, representing a
conducting cathode material which is strictly correct only for
$T>$1800K from which iron dominates the grain material composition in
the cloud. As mentioned earlier, it is only relevant to consider the
material dependence of $\gamma$ for the avalanche process, not so for
the streamer.

Figure ~\ref{fig:capdist} shows that the breakdown distance varies
over the atmosphere pressure ranges from $\sim 5\mu$m in the densest
regions to $10^5$km in the upper, low pressure regions (dashed lines).
This suggests that discharges could appear spatially extended and
diffuse in low-pressure regions, while numerous discharge events
remain small, locally confined phenomena in high-pressure regions.  It
also suggests that different electric discharge processes dominate at
different heights: local corona (point discharges) and small-scale
sparks at the bottom of the cloud, and glow discharge and large-scale
sparks at the top.

In a quasi-static environment, these different processes could lead to
a neutralisation of large-scale charge built-up and consequently
diminish the possibility for a large-scale lightning. However,
hydrodynamic atmosphere  motions are driven by convection and by gravity waves
(Freytag et al. 2010) on Brown Dwarfs which are also rapid rotators
(Scholz et al. 2011).  This suggests that large-scale charge
separation is likely to be sustained by convection-like processes
comparable to Earth, for example like in the Hadley cells at the
Earth's equator.

Following the observation in Raizer (1991) that for breakdown
distances $>5\,\ldots\, 6$cm the streamer mechanism will dominate and
sparks can develop, we distinguish a large-scale (d$> 6$cm) and a
small-scale regime (d$< 6$cm) in
Fig~\ref{fig:capdist}. 

It is interesting to compare the maximum spacial separation that two
charge distributions whilst still permitting a discharge, to the
average size of the cloud particles and their mean particle
distance. For simplicity, we consider the altitude dependent mean
grain size $\langle a \rangle$ instead of separate grain size
distributions for each altitude as shown in Helling, Woitke \& Thi
(2008).  The mean grain size, $\langle a \rangle$, is always smaller
than the maximum distance, $d$, over the whole geometrical extension
of the cloud. The mean distance between the cloud particles scales
with the cloud particle number density as $n_{\rm
  d}^{-1/3}$. Consequently, it will be largest where the number
density is smallest. The 'knee' at the high-pressure end of each
long-dashed curve in Fig.~\ref{fig:capdist} corresponds to the maximum
cloud particle number density (compare the upper panel in
Fig.~\ref{fig:fracide}). Generally, the cloud particles have a mean
separation that is $\sim$ 3.5 orders of magnitude smaller than the
maximum break-through distance in a giant gas planet cloud ($\sim$ 2.5
for the Brown Dwarf cloud). This comparison also shows that the cloud
particles have considerable separation in the upper cloud region
suggesting that collective phenomena such as large-scale charge
distributions are to be expected rather than inter-particle discharges
such as in the high-pressure part of the cloud. However, if the
breakdown distance is smaller than the maximum value $d$ (e.g. in a
turbulent flow), small-scale discharge processes may also occur in the
low-pressure part of an atmosphere.

Note, the mean particle size and the cloud particle number density are
determined by the processes of dust and cloud formation (nucleation,
growth/evaporation, drift, convective mixing, element depletion) taken
into account in the {\sc Drift-Phoenix} model
atmospheres. Coagulation, for instance, could change the grain size
distribution, but it is not clear {\it a priori} how it changes since a
variety of constructive and destructive collisional processes occur
(G\"uttler et al. 2010). Furthermore, turbulence (Helling et al. 2001,
2004) and gravity waves (Freytag et al. 2010) cause local,
hydrodynamic density increases, neither of which is taken into account
in the present paper.

Depending on the ionising gas, these global values of $d$ have
uncertainties of one order of magnitude as we demonstrate for
different gas-phase compositions in Fig.~\ref{fig:capdist_dg}. The
breakdown distances inside the cloud regions (solid lines,
Fig.~\ref{fig:capdist}) are somewhat different in a Brown Dwarf
mineral cloud and in a giant planet's mineral cloud because the Brown
Dwarf clouds exist at higher pressures. The exact values of
discharge quantities, such as the maximum separation of charge
distributions before breakdown but also the required number of charges,
will therefore differ between objects of different effective
temperature, surface gravity and metallicity (see
Figs.~\ref{fig:Qcrit_Teff}, ~\ref{fig:Qcrit_metal}).

\paragraph{Minimum electric field for breakdown $E_{\rm t, min}$:}
The minimum electric field strength, needed for a gas breakdown scales
with the local gas pressure (Eq.~\ref{eq:paschenmin_Ecrmin}), are shown
in Fig.~\ref{fig:Etmin}. This critical breakdown field strength
corresponds to the minimum of the Paschen curve for individual ionised
gases (H$_2$, He, air, H$_2$O, CO$_2$; Table 4.1 in Raizer 1991)
represented by individual parameters B [V cm$^{-1}$ torr$^{-1}$], and
a given Townsend's 2nd ionisation coefficient $\gamma=0.001$.  The
results are shown for a sample model atmosphere (T$_{\rm eff}$=1600K,
log(g)=5, initial solar composition) for which we vary the parameter B to show the
dependence on gas composition. Figure~\ref{fig:Etmin} shows results
for pure gases (data from Raizer 1991), and Fig.~\ref{fig:Etmin_zoom}
for mixed gases of the solar system planets (data from Sentman 2004).
Over-plotted in Fig.~\ref{fig:Etmin} are the values given in Yair et
al. (1995: magenta - Jupiter cloud with drops and ice, cyan - Earth
values; full symbol - Paschen formula, open symbols - measurements).
This shows that the values are comparable.  The vertical lines in
Fig.~\ref{fig:Etmin} indicate the large-scale regime (low pressure)
and the small-scale regime (high pressure) for pressures corresponding
to Fig.~\ref{fig:capdist_dg}.

 The minimum critical breakdown field strength  varies between $10^7$ V/cm at high atmospheric
pressures and $10^{-7}$ V/cm in low pressure regions where $p_{\rm
  gas}=10^{-12}$ bar. The breakdown fields on Earth (at sea level) and in
Jupiter are $\sim 10^4\,\ldots\,10^6$ V/cm which correspond to the
values at 1 bar in Figs.~\ref{fig:Etmin} and \ref{fig:Etmin_zoom}. The
dependence on the local gas composition introduces an uncertainty of
one order of magnitude as demonstrated in Fig.~\ref{fig:Etmin}.


Figure~\ref{fig:Etmin_zoom} shows a close-up of Fig.~\ref{fig:Etmin}
showing the minimum breakdown field using the data for the
different solar system planets in comparison with the two most
different examples of gas phase composition (CO$_2$ and He gas). The
results for the solar system planets sit between 
a pure H$_2$O (upper limit) and a pure H$_2$ atmosphere
(lower limit).

 We note further that values measured for the onset of coronal
  discharges on a treeless plot of grass are about 3 orders of
  magnitude lower (40 V/cm; MacGorman \& Rust 1998) than the
  respective value at 1 bar in Figs.~\ref{fig:Etmin},
  \ref{fig:Etmin_zoom}. MacGorman \& Rust (1998) also point out that
  point discharges from ice crystals which are dominated by their
  surface (rather than bulk) conductivity, seem likely in thunderstorms for
  a field strength $> 400$ kV/m ($4\cdot 10^3$ V/cm) which is still
  $\sim 2$ orders of magnitude lower than the classical values
  in Figs.~\ref{fig:Etmin} and \ref{fig:Etmin_zoom}.

Therefore, our results suggest that electrical breakdown can readily
occur in cloudy atmospheres of Brown Dwarfs and giant gas planets.

\subsection{How many charges are needed  for an electric field breakdown?}\label{ss:hmc}

Laboratory measurements of grain charges suggest that single grains
can carry $10^3-10^5$ e/grain. This is supported by experiments on
photoelectrical charging of dust in a vacuum (Sickafoose et al. 2000;
metals and glass), grain levitation (Fortov et al. 2001),  measurements of
 grauple on Earth (Lamb \& Verlinde 2011), and wind-driven entrainment of
Martian dust (Merrison et al. 2012; silicates and iron
oxide). Somewhat higher charge loads are suggested from dust-dust
collision experiments (e.g. contact electrification and tribo-electrification)
with silicate granules by Poppe, Blum \& Henning (2000) resulting in
$10^{-5}$C/m$^{-2}$ (i.e., $\sim 5\cdot10^6$e/grain for a=1.2$\mu$m),
and by volcanic ash experiments involving dust charging by
fractoemisson by James et al. (2000) resulting in
$10^{-8}\,\ldots\,10^{-6.5}$C/m$^{-2}$ ($\sim
10^5\,\ldots\,10^{7.5}$e/cm$^2$ ). How do these experimental numbers
compare to the number of charges needed to initiate a field breakdown
in a substellar atmosphere?

The minimum electric field strength that needs to be overcome in order
to produce a field breakdown in an astrophysical gas with a low
degree of ionisation (see Fig.{~\ref{fig:fracide}, upper panel, solid
  lines) varies only by one order of magnitude between gases of
  different molecular composition for a given gas
  pressure. Furthermore, the breakdown voltage measured above
  thunderclouds on Earth is up to two order of magnitude lower than
  the classical breakdown values that were discussed here so far 
    (see also the discussion at the end of Sect.~\ref{ssBDQs}). Keeping these
  uncertainties in mind, we continue with studying how many charges
  are needed to initiate a field breakdown in a gas of a given
  composition for the minimum breakdown voltage $V_{\rm t, min}$.

The assumption of a spherically symmetric charge distribution
(Eq.~\ref{eq:Qcrit}) allows a first order-of-magnitude study, although
neither clouds nor cloud particles are truly
spherical. Equation~\ref{eq:Qcrit} is a general expression which allows
us to distinguish between these two scale regimes of clouds and grains
later on. With the radial distance from a charge distribution of
$Q_{\rm crit}$ charges set to the minimum distance for breakdown,
$d$, it follows that
\begin{eqnarray}
\label{eq:Qcrit}
Q_{\rm crit}&=& 4\pi \epsilon_0\,d\,V_{\rm t, min}  \,\, \mbox{[C]} \qquad\,\,\Rightarrow Q_{\rm crit} \propto d \propto \frac{1}{p\mbox[\rm Pa]}.
\end{eqnarray}
1 electron carries only the tiny fraction of a Coulomb: 1C = $6.24150965\,10^{18}$ e.
  The surface charge density for a charge distribution of radius $r$
(grain radius or cloud radius) is given by $\sigma=Q/(4 \pi
r^2)$. Thus, the critical surface charge density  for
breakdown is
\begin{eqnarray}
\label{eq:sigcrit}
\sigma_{\rm crit} &=& \frac{\epsilon_0\, d\, V_{\rm t, min}}{r^2} = \epsilon_0 E_{\rm r, min} \Big(\frac{d}{r}\Big)^2\,\,\,\ \mbox{[C/cm$^2$]} 
\end{eqnarray}
($\epsilon_0=8.85\,10^{-12}$F\,m$^{-1}$(\footnote{$\epsilon_0$ is in
SI unites F\,m$^{-1}$ with F=As/V (A - Ampere) = C/V (C-Coulomb) =
C$^2$/(Nm)}) - vacuum dielectric constant or electric permittivity of
free space).  Note that the proportionality at the right hand side of
Eq~\ref{eq:Qcrit} {\it only holds if} the radial distance from the
charge distribution is set to the minimum distance for breakdown
according to the minimum of the Paschen curve
(Eq.~\ref{eq:paschenmin_pd}). Note further that the pressure needs to
be in SI units [Pa]  for $Q_{\rm crit}$ to be in units of elementary
charges on the right hand side of Eq.~\ref{eq:Qcrit}.

The charge carrying surface, $4 \pi r^2$, of our charge distribution
can be the effective surface of a cloud or the surface of a single
grain ($r=a$).  The total charge carrying surface of the cloud, the
effective cloud surface, will be provided by the total surface of the
particles that form the cloud, rather than being related to the
geometrical cloud radius. On large vertical scales, the cloud
particles will settle depending on their masses and sizes. Our
stationary cloud formation model suggests a large mean grain size at
the lower edge of the clouds and a small mean grain size in the
cloud's top region (Woitke \& Helling 2004; Witte, Helling \&
Hauschildt 2009). Smaller particles will remain suspended for longer
in the atmosphere than the more positively charged bigger grains. On
small scales, the charge separation needed to build up an electric
field is influenced by non-spherical grain shapes and chemical surface
inhomogeneities, neither of which is included in our considerations
yet. For spheroidal grain growth, we refer to Diver \& Clark (1997)
and Stark et al. (2006).

Our cloud formation model (Woitke \& Helling 2003, 2004; Helling
\& Woitke 2006; Helling, Woitke \& Thi 2008) allows us to link
 numerically derived cloud properties, like  grain
sizes, grain number densities, and dust surface, to critical breakdown quantities as
evaluated in the previous sections. The solution of the our cloud
model in the frame of the radiative transfer code {\sc Phoenix}
(Hauschildt \& Baron 1999) allows to link dust properties and the
critical breakdown quantities to the atmospheric temperature and
pressure scale of a Brown Dwarf or giant gas planet as result of the
{\sc Drift-Phoenix} model atmosphere simulations (Dehn 2007, Helling
et al. 2007a,b; Witte, Helling \& Hauschildt 2009).

The moment equations of our kinetic dust formation model (Gail \&
Sedlmayr 1988, Dominik et al. 1993, Woitke \& Helling 2003) allow us
to calculate the total dust surface available per unit volume ($A^{\rm
tot}_{\rm dust}$), and also the mean surface of a grain ($<\!\!A_{\rm
dust}\!\!>$) depending on atmospheric height (i.e. depending on the local temperature and pressure),
\begin{eqnarray}
\label{eq:Atot}
A^{\rm tot}_{\rm dust} &=& \sqrt[3]{36\pi}\, L_2  \,\,\,\,\mbox{[cm$^2$/cm$^3$]}\\
\label{eq:Amean}
<\!\!A_{\rm dust}\!\!> &=&
\sqrt[3]{36\pi}\, L_2/L_0 = A^{\rm tot}_{\rm dust}/n_d \,\,\,\,\mbox{[cm$^2$]}.
\end{eqnarray}
 $L_0$ is the zeroth and $L_2$ the second moment of the grain size
distribution function, respectively (e.g. Helling et
al. 2001). Equations~\ref{eq:Atot} and \ref{eq:Amean} allow us to relate
the total number of charges needed for breakdown to the total grain
surface per unit volume (upper panel, Fig.~\ref{fig:QcritxA}) and to
the numbers of grains (lower panel, Fig.~\ref{fig:QcritxA}) available
as follows:
\begin{eqnarray}
\sigma\times  A^{\rm tot}_{\rm dust} \,\,\mbox{[e$^-$/cm$^3$]\,\,} &-&  \mbox{number of charges per total dust surface
 per  gas volume}\\
\sigma\times  \frac{A^{\rm tot}_{\rm dust}}{n_d}\,\,\mbox{[e$^-$/grain]}\,\, &-& \mbox{number of charges per grain.}
\end{eqnarray}
 Note that the total dust surface, i.e. the total cloud particle surface, is
 related to the charge carrying surface inside a cloud.

The number of charges required to establish an electric field large
enough for a breakdown to occur and subsequently to ionise the ambient
gas, depends on the local gas pressure. More charges are needed in the
low density (pressure) regime of an atmosphere as for example
Fig.~\ref{fig:Qcrit} demonstrates because of the large maximum
breakdown distance (compare Figs.~\ref{fig:capdist},
\ref{fig:capdist_dg}). The lowest density layers would require a total
of about $10^{20}$ charges to allow an electron avalanche with a
subsequent streamer to develop. The actual number depends to a certain
extent on the dominating gas species that is being ionised as in the
example for H$_2$O and CO$_2$ in Fig.~\ref{fig:Qcrit}. The lower panel
of Fig.~\ref{fig:Qcrit} presents this results in (SI) units of
critical charge density, $\sigma_{\rm crit}$ [C/cm$^2$], assuming that
the charges would be located on the grains of the size of the local
mean grain size, hence $r=<\!\!a(T,\rho_{\rm gas})\!\!>$.
Artificially increasing the radius of the charge carrying surface to
$r=100\times\!\!<\!\!a(T,\rho_{\rm gas})\!\!>$ (long dashed red line)
verifies the anticipated result of a decreasing critical charge
density with increasing surface area that would allow an electric
field breakdown.

Figure~\ref{fig:QcritxA} (upper panel) shows the critical number of
charges, $\sigma_{\rm crit}\times A^{\rm tot}_{\rm dust}$
[e$^-$/cm$^3$], as distributed over the total dust surface per cm$^3$
of atmospheric gas depending on altitude (represented by the local
atmospheric gas pressure). The dust surface is provided by a large
number of large dust grains at higher pressure and by small number of
smaller grains at lower pressures at the cloud top.  If the same total
dust surface would be provided by grains of a constant size of
$10^{-3}$cm (brown long-dash short dash line), the critical number of
charges per cm$^3$ would decrease by more than 5 orders of magnitude.

The lower panel of Fig.~\ref{fig:QcritxA} shows the critical number of
charges per grain, $\sigma_{\rm crit}\times \frac{A^{\rm tot}_{\rm
    dust}}{n_d}$ [e$^-$/grain], which takes into account the total
dust surface and the number of grains present throughout the
cloud. Each grain has to carry considerably less charge at the bottom
of the cloud compared to the cloud top as here more grains are
available. Due to gravitational settling, most of the grains are
contained in the deeper cloud where the number of grains is much
higher than at the cloud top.

Our results suggest that many charges per grain are required to
initiate a breakdown discharge in the upper part of the clouds in
Brown Dwarfs and giant gas planets.  Such a height-dependent charge
distribution results because we apply our results for the maximum
breakdown distance, $d$, which is rather large at the cloud top as
demonstrated in Figs.~\ref{fig:capdist}, \ref{fig:capdist_dg}. The
breakdown distance can, however, decrease by e.g. turbulent or other
non-local wind processes which are not taken into account here. A
decreasing number of charges needed to initiate an electric field
breakdown would then result.  Additionally, the charge density,
$\sigma$, depends on the radius of the charge carrying surface,
$r$. Therefore, we again assess our results for $r$ larger than
$<\!\!a\!\!>$, but now by using a constant grain size of $10^{-3}$cm
(brown long-dash short dash line) throughout the cloud. The comparison
with the experiments with mono-disperse grain ensembles as performed
by Sickafoose et al. (2000; green star in Fig.~\ref{fig:QcritxA}),
Fortov et al. (2001; magenta vertical bar in Fig.~\ref{fig:QcritxA}),
Merrison et al. (2012; blue box in Fig.~\ref{fig:QcritxA}), Poppe,
Blume \& Henning (2000; blue asterisk in Fig.~\ref{fig:QcritxA}),
James et al (2000; blue asterisk in Fig.~\ref{fig:QcritxA}), and
  values for grauple on Earth from Lamb \& Verlinde (2011; orange
  vertical bar in Fig.~\ref{fig:QcritxA}) suggest that our results
for mineral clouds are not unreasonable, especially if we allow that
cloud particle sizes can be larger than the mean grain size from our
dust model.  Such a deviation from the mean grain size is to be
expected as our study of the grain size distribution in Helling,
Woitke \& Thi (2008) demonstrates. Furthermore, critical
breakdown field strengths applied here  are upper values as experiment
on Earth thunderstorms suggest and can, hence, be expected to be 1-2
orders of magnitude lower with the corresponding decreasing effect on
the necessary number of charges.  The chemical richness of the
  mineral clouds in the atmosphere of Brown Dwarfs and giant gas
  planets may introduce further uncertainties compared to the
  laboratory experiments cited above. The actual material composition
  may influence the charge distribution on the grain surface leading, for example, to
further polarisation effects of the cloud particles.  The question of
how many charges a dust grain can carry before it is destroyed by the
electrostatic stress will be addressed in a forthcoming paper.

\paragraph{Trends with T$_{\rm eff}$ and metallicity [M/H]:}
The quantities that determine if the streamer mechanism occurs and
with which efficiency it does so, are determined by the local gas
pressure because the gas pressure determines the mean free path of the
seed electrons inside the gas of Brown Dwarfs and planets in the
electric field. Global quantities that determine an atmosphere's
pressure structure are the effective temperature, T$_{\rm eff}$, the
metallicity, [M/H], and the surface gravity, log(g), of the object,
in the case of non-irradiated objects.  The influence of the surface
gravity has already been shown in Figs.~\ref{fig:Qcrit} and
~\ref{fig:QcritxA}. We use the surface gravity to distinguish between
Brown Dwarfs (log(g)=5.0) and giant gas planets (log(g)=3.0).

Figures~\ref{fig:Qcrit_Teff} and ~\ref{fig:Qcrit_metal} show the
influence of T$_{\rm eff}$ and [M/H], respectively, for a small number of {\sc Drift-Phoenix} models.  The metallicity
[M/H] is given relative to the solar abundances, hence, [M/H]=$-$3.0
stands for the initial element abundances being decreased by
$10^{-3}$ compared to the solar values.  The lowest panels in both
figures demonstrate how the {\sc Drift-Phoenix} temperature-pressure
profiles change with changing metallicity. The local pressure mainly
increases for a given local temperature for an increasing T$_{\rm eff}$
(Figures~\ref{fig:Qcrit_Teff}), while the upper atmospheric parts of
the (T$_{\rm gas}$, p$_{\rm gas}$)-structures stay relatively
comparable with decreasing metallicity for given T$_{\rm eff}$ and
log(g). The inner atmospheric regions, however, show an increasing gas
pressure with decreasing metallicity for a given gas temperature.

The critical charge number density needed to initiate the occurrence
of streamers that subsequently might lead to lightning, does change by
up to 4 orders of magnitude for a given gas pressure amongst the
models of varying effective temperature that form dust clouds. Note
that the T$_{\rm eff}=2800$K-model in Fig.~\ref{fig:Qcrit_Teff} is too
hot for dust condensation processes to be efficient. The metallicity
of the atmospheric gas has a larger impact on the local
temperature. Figure~\ref{fig:Qcrit_metal} shows that the critical
charge number density can differ in the low-metallicity cases much
more than for varying T$_{\rm eff}$. Note that the dust surface
available to host the charges does vary amongst the models, with the
largest differences occurring again for changing metallicities in the
lowest metallicity cases (dashed/dotted line) compared to the solar
case (solid line). All models, however, exhibit a comparable total
dust surface in the inner, denser regions at the bottom of the cloud.
Witte et al. (2009) demonstrated that low-metallicity atmospheres can
produce larger dust grains at the inner edge of the cloud than in the
solar case if [M/H]$>=$-4.0.

 \subsection{Number of electrons vs. number of grains}\label{sec:efficient}

So far, we have studied the breakdown conditions utilising classical
arguments and laboratory measurements assuming that the gaseous
atmosphere would have enough free electrons available for starting the
electric breakdown process.  But, is the number of thermal electrons large
enough in Brown Dwarf and giant gas planet atmospheres to provide the
seed electrons that are needed to start an electron avalanche in the
electric field, for example between two passing grains? The
investigation of this question allows us to estimate the efficiency
with which streamer discharges could occur inside Brown Dwarf and
giant gas planet atmospheres. Streamer discharges are the starting
point for large-scale phenomena like lightning  that may be
  followed by a sprite into the less dense atmosphere above.

We compare the degree of thermal ionisation, $\chi_{\rm e,
themal}=n_{\rm e, thermal}/n_{\rm H}$ (solid lines), with the number
of dust grains per hydrogen, $n_{\rm d}/n_{\rm H}$ (dashed lines) in
Fig.~\ref{fig:fracide} (top panel), and we use these two quantities to
estimate an efficiency of streamer initiation in a mineral cloud by
calculating the ratio $n_{\rm e, thermal}/n_{\rm d}$
(Fig.~\ref{fig:fracide}, bottom panel). Hence, this efficiency would
be the number of streamers per grain pair if each of the thermal
electrons in the electric field of two charged grains would initiate an
electron avalanche. In the previous section, we have evaluated the
conditions under which such a field breakdown can occur.

Figure~\ref{fig:fracide} shows that the efficiency, $n_{\rm e,
thermal}/n_{\rm d}$, varies with height as at some height the dust
number density can have a local maximum while the number of thermal
electrons is a rather smooth function of gas pressure which is linked
to the gas temperature by the underlying atmosphere model. Hence, the
number of electrons available for each colliding pair of
charge-carrying grains decreases as shown in the lower panel of
Fig.~\ref{fig:fracide}. Here, more than one electron would be
available for initiating the avalanche process for all pressures above
the solid black, horizontal lines ($n_{\rm e,thermal}= n_{\rm d}$),
and less than one electron would be available below this line. In
both cases, the Brown Dwarf atmosphere (brown) and the giant gas
planet's atmosphere (red), the efficiency of streamer initiation would
vary between 1 and $\approx 10^{-5}$.  $n_{\rm e,thermal}/ n_{\rm
d}=1$ means that one streamer per grain pair sets off, and $n_{\rm
e,thermal} < n_{\rm d}$ means that not every grain pair can initiate a
streamer event. To generalise, we can distinguish two regimes regarding
the streamer efficiency in a cloud forming atmosphere, the occurrence of
which can be less homogeneous if the whole grain size distributions for each
atmospheric layer are taken into account,
\begin{tabbing}
$n_{\rm e,thermal} > n_{\rm d}\qquad$ \= electron dominated$\qquad$ \= $n_{\rm e,thermal}/ n_{\rm d}\,\, \gg 1$\\
$n_{\rm e,thermal} < n_{\rm d}\qquad$ \= dust dominated   \> $n_{\rm e,thermal}/ n_{\rm d}\,\,\leq 1\,\ldots\,10^{-5}$.
\end{tabbing}

The consequence is that the discharge process in cloudy atmospheres
will be determined by  two populations, the population of the cloud
particles and the population of free electrons, respectively. Both
populations need to be abundant enough to allow discharge processes to
develop into streamers as the pre-condition for the occurrence of  lightning.

 \subsection{An estimate of electron enrichment of the gas phase by dust-dust induced streamer events in mineral clouds}\label{ss:disc}

\noindent
The empirical formula
\begin{equation}
\label{eq:townsend}
\alpha = A\,p \exp\big(\frac{-Bp}{E}\big) \quad \Rightarrow \quad \alpha(E_{\rm min})=\frac{Ap}{e}
\end{equation}
for Townsend's ionisation coefficient, $\alpha$ [cm$^{-1}$], enables
us to provide first estimates of the number of electrons created when a
electron avalanche starts that might develop into a streamer.   Streamers are discussed to be a likely possibility to initiate
  lightning (see Sect. 5.4. in MacGourman \& Rust 1998). $\alpha$ is the
number of ionisation events performed by an electron in a 1cm path
along an electric field.  This interpretation is only true if we
assume that the electron undergoes ionisation collisions only for high
values of $E/p$ and moderate electron energies (Raizer 1991). The
constants $A$ and $B$ are the same constant given in
Table~\ref{tab:AB} for approximating experimental curves.  $\alpha d$
would then be an estimate for the number of electrons produced if the
electron avalanche would stretch along the whole critical breakdown
distance $d$ (compare Fig.~\ref{fig:capdist}). The total number of
non-thermal electrons per unit time, $N^{\rm aval}_{e} [s^{-1}]$, produced
by all possible seed electrons that lead to an electron avalanche, can
be approximated by
\begin{eqnarray}
\label{eq:Neaval}
N_{\rm e}^{\rm aval} &\approx& \,\,\,\,\,f_{\rm eff} \,\,\,\,\times \alpha d \times \nu_{\rm coll}^{dd}(z)\\
                         &\approx& \frac{n_{\rm e, thermal}}{1/2\cdot n_{\rm d}} \times \alpha d \times \nu_{\rm coll}^{dd}(z),
                         \end{eqnarray}
with $ \nu_{\rm coll}^{dd}(z)$ the frequency of dust-dust collisions
at each height $z$ in the atmosphere as derived in Eq.17 in Helling,
Jardine \& Mokler (2011). Following from Sect.~\ref{sec:efficient}, we
define $f_{\rm eff} = n_{\rm e, thermal} / (1/2\cdot n_{\rm d})$ which
provides an estimate of the efficiency with which an avalanche could
occur. $f_{\rm eff}>1$ suggests a superposition of avalanches that
are each started by one gas-phase electron.

An electron avalanche can turn into a self-propagating ionisation
front by the subsequent development of a streamer in its positively
charged wake. This instability can  trigger much larger plasma-channel
structures like sprites above thunderclouds or lightning bolts below
and inside a cloud which we are not considering for our estimate here.
The elementary streamer alone, however, results in a number of local
non-thermal electrons much larger than predicted by the classical
Townsend breakdown by electron avalanche
(Eq.~\ref{eq:townsend}). Numerical experiments suggest the production
of $10^{12}\,\ldots\,10^{13}$ electrons per streamer event (e.g. Dowds
et al. 1993). Multiple streamers develop if more than one electron is
'seen' by a pair of charged grains or two otherwise oppositely charged
charge distributions. This efficiency depends on the number of free
electrons that are already present in the gas before a
streamer-triggering electron avalanche develops and has been derived
in Sect.~\ref{sec:efficient}.  For a streamer breakdown, the number
of non-thermal electrons produced during a collisional approach of two
dust particles that occur with the frequency $\nu_{\rm coll}^{dd}(z)$,
\begin{eqnarray}
\label{eq:Nestre}
N_{\rm e}^{\rm streamer}  &=& \frac{n_{\rm e, thermal}}{1/2\cdot n_{\rm d}} \times  10^{12} \times \nu_{\rm coll}^{dd}(z).
                         \end{eqnarray}

Given the different nature of an electron avalanche and a streamer,
the electron production occurs with a different efficiency. The
result is that $N_{\rm e}^{\rm streamer}\gg N_{\rm e}^{\rm aval}$
(Fig.~\ref{fig:Nestream}).  Our results in Fig.~\ref{fig:Nestream}
suggest further that the largest impact of the streamer electrons on the
local chemistry could be expected in the inner, denser cloud regions
in Brown Dwarfs and giant gas planets.


\section{Conclusions}\label{sec:concl}

Discharge process require a) charge separation on cloud particles and
b) a charge separation over large distances to allow large-scale
discharges such as lightning or sprites to occur.  The laboratory
experiments (Poppe, Blum \& Henning 2000, James et al. 2000, Fortov et
al. 2001, Lacks \& Levandovsky 2007, Forward et al. 2009, Merrison et
al. 2012) for a variety of minerals (including Martian and volcanic
minerals),  for graupel on Earth (Lamb \& Verlinde 2011), as well as our own investigations (Helling, Jardine, Mokler
2011) suggest that mineral cloud particles will be charged. Mineral
cloud particles settle gravitationally in Brown Dwarfs and giant gas
planets such that a large scale charge separation can occur inside
these clouds.  Convection serves as an additional large-scale
  charge separation mechanism. If small grains are indeed more
negatively charged than large grains as suggested by experiments
(Hatakeyama \& Uchikawa 1952, Kikuchi \& Endoh 1982, James et
al. 2000, Merrison et al. 2012), spatial charge separation will occur
on a large scale by gravitational settling but also on small scales by
particle size effects alone.  Our investigations of the minimum
distance that two charge distributions need to have for a electric
field breakdown to occur (Fig.~\ref{fig:capdist}) suggest that
small-scale discharges should be expected in high-pressure regions of
an atmospheric cloud. Large-scale discharge processes should occur in
the upper, low-pressure part of the cloud. This is consistent with
observations from Earth where, for example,  sprites develop into the low-pressure
part of the atmosphere above a thunder-cloud event.

The critical electric field strength to be overcome for a field breakdown varies between $10^7$ V/cm at high atmospheric pressures and $10^{-7}$ V/cm in the upper atmosphere where the gas pressure is low.
The critical number of charges per total dust surface  per cm$^3$ change
from $10^{23}$ e/cm$^3$ in the inner atmosphere to  $<10^{5}$ e/cm$^3$ in the low pressure atmosphere. These numbers compare well with experimental values at the 1bar-pressure level.
The number of charges  needed to initiate the occurrence of
streamers that subsequently might lead to lightning, does not change
substantially amongst the atmosphere models of varying effective temperature that
form dust clouds. The critical number of charges has the same trend in
the low-metallicity cases as for varying effective
temperatures.

Generally, a population of charged cloud particles and enough free
electrons are needed to allow transient luminous events to occur in
extraterrestrial, cloudy atmospheres. The ratio of both determines the
efficiency with which discharge events may occur. The rate of electron
enrichment is highest in the high-pressure part of the cloud as here,
according to our model, a larger charge-carrying surface is available. This is where we expect  small-scale 
discharges to dominate.



{\bf Acknowledgement:} 
We highlight financial support of the European Community under the FP7
by an ERC starting grant. Y. Yair is thanked for providing Sentman (2004).  ChH acknowledges the hospitality of the University of Vienna during the beginning of the paper's work.
Most literature search was performed using the ADS. Our local computer support is highly acknowledged.

\label{lastpage}

\clearpage



\begin{figure}
\hspace*{1.3cm}\resizebox{10.5cm}{!}{\includegraphics{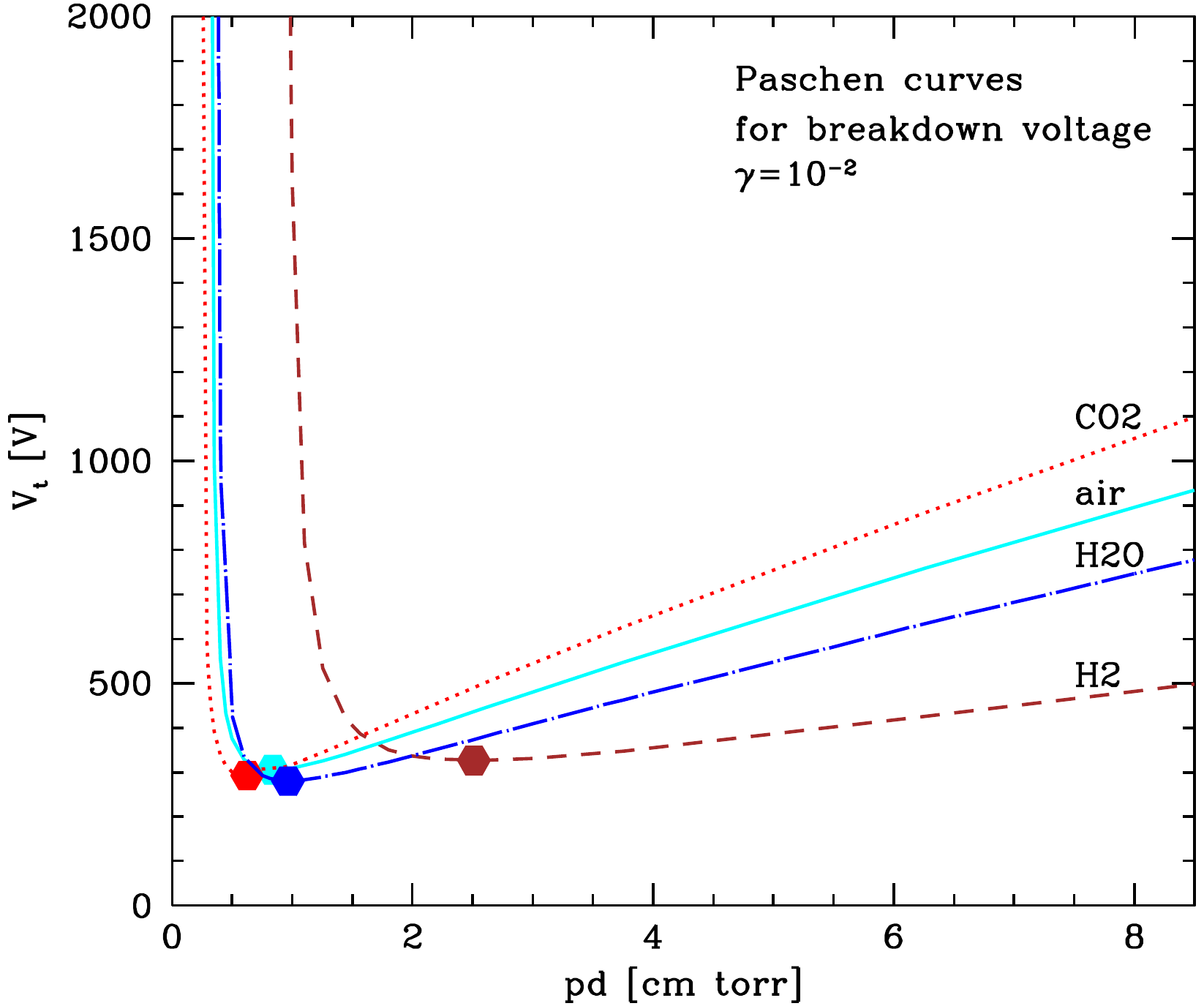}}
\caption{\small The Paschen curve shows the dependence of the
breakdown voltage, $V_{\rm t}$ [V] on the product of gas pressure,
$p$, and the minimum breakdown distance, $d$, for four molecular
gases composed of H$_2$, H$_2$O, CO$_2$, and of air. The classical
Paschen curve depicted here, has a minimum (indicated by the symbol;
Stoletow point) which we use to determine the minimum electric
breakdown field  in Brown Dwarf and exoplanetary atmospheres in
Figs.~\ref{fig:capdist}-~\ref{fig:capdist_dg} }
\label{fig:paschen}
\end{figure}

\begin{figure}
\hspace*{1.3cm}\resizebox{10.5cm}{!}{\includegraphics{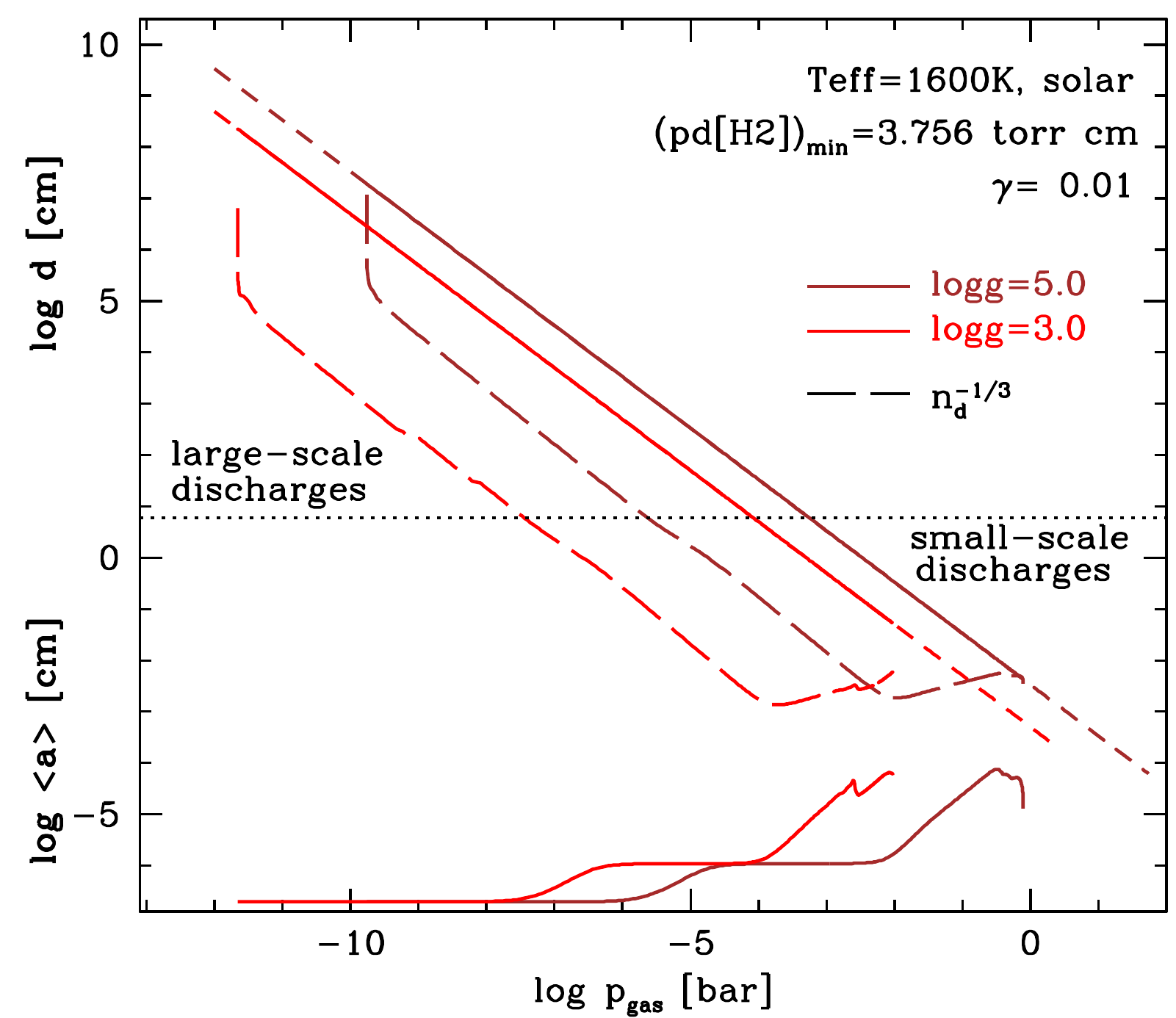}}
\caption{\small The minimum separation of charges of two capacitor
plates (representing our charge carrying surfaces like on clouds or on
grains) required for discharge (solid and short-dashed in the upper
part of the plot). The distance $d$ corresponds to the Paschen minimum where $p_{\rm
gas}\cdot d= 2.509$ torr cm (=5007 dyn/cm$^2$ cm) for an H$_2$
gas. Note, an  artificial offset between the log(g)=5.0 (brown, "logg=5.0") and the log(g)=3.0
(red, "logg=3.0") was applied; all curves should lie on top of each other without
this offset. $p_{\rm gas}\cdot d$ is a constant for a given $A$
[cm$^{-1}$ torr$^{-1}$] and $\gamma$, hence, the same for all
atmospheres of the same chemical composition.  In this plot, an
H$_2$-gas is assumed for which A[H$_2$]=5 cm$^{-1}$Torr$^{-1}$ (Raizer
1991) and Townsend's 2nd ionisation coefficient $\gamma=0.001$ is
used. The solid lines show the pressure interval where clouds form and
the dashed lines show the whole atmospheric pressure range modelled.
Raizer (1991) notes that discharge sparks set in if the plate distance
is $>$6cm (horizontal dotted line) which helps us here to distinguish
between large-scale and small-scale discharges. We overplot two more
length scales for comparison: The dust-dust mean free path $n_{\rm d}^{-1/3}$ (long
dashed lines) and the mean grain sizes, $\langle a \rangle$, at the
bottom of the plot.}
\label{fig:capdist}
\end{figure}

\begin{figure}
\hspace*{1.3cm}\resizebox{10.5cm}{!}{\includegraphics{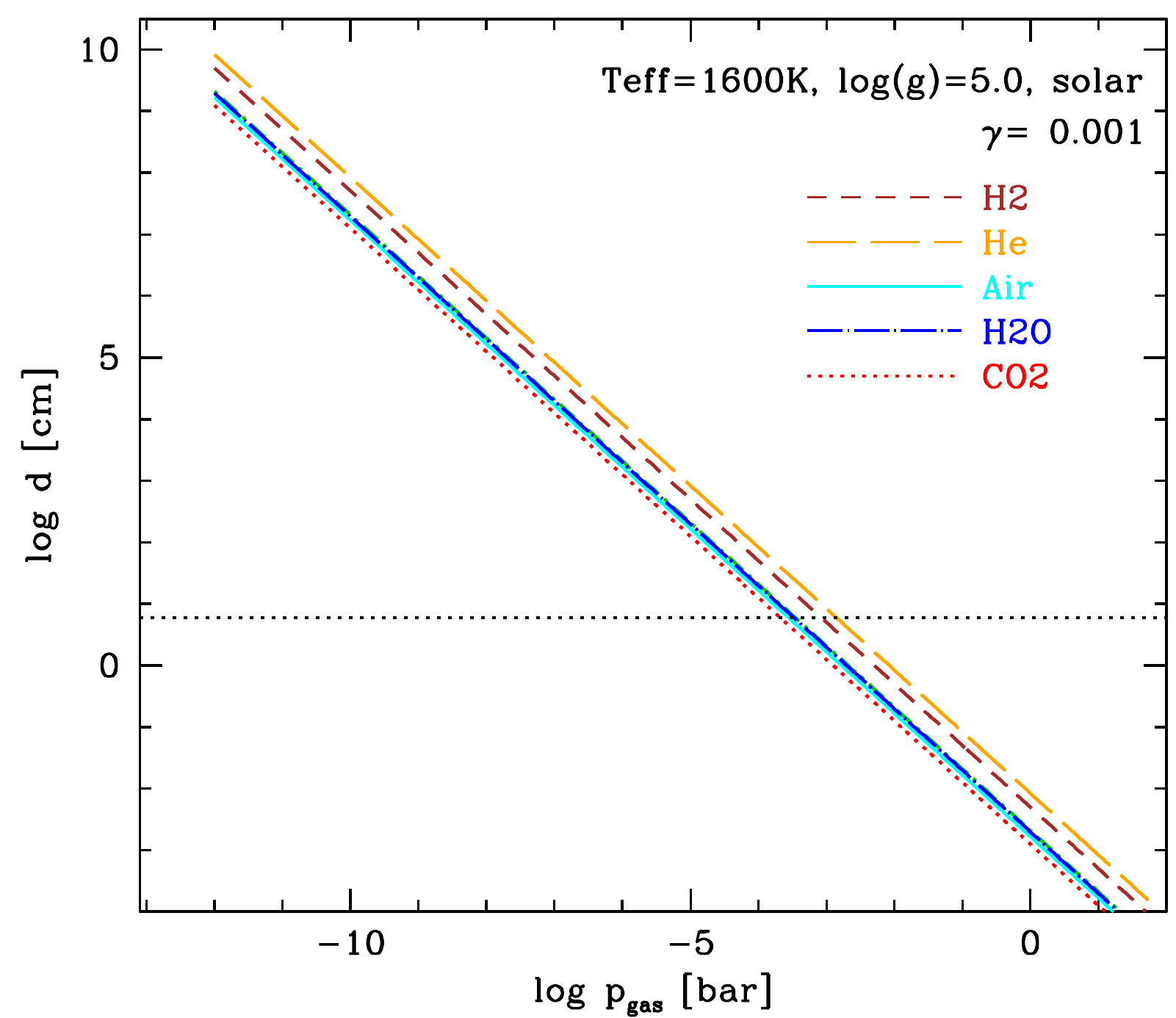}}
\caption{\small Dependance of breakdown distance $d$ at the Stoletow point
on the composition of the ionised gas. The variation in these distances for 6
cases (H$_2$, He, air, H$_2$O, CO$_2$; Table 4.1 in Raizer 1991) shows
that the uncertainty resulting from atmospheric composition is a
factor of 10. This estimate also covers differences in $A$ depending
on the pressure regime in which the measurements were made (high
[$\sim 1$ bar] vs. low pressure; Table 7.1 in Raizer 1991). Sentman
(2004) has published data for mixed gases but the lack of a consistent notation
makes application difficult.}
\label{fig:capdist_dg}
\end{figure}

\begin{figure}
\hspace*{1.3cm}\resizebox{10.5cm}{!}{\includegraphics{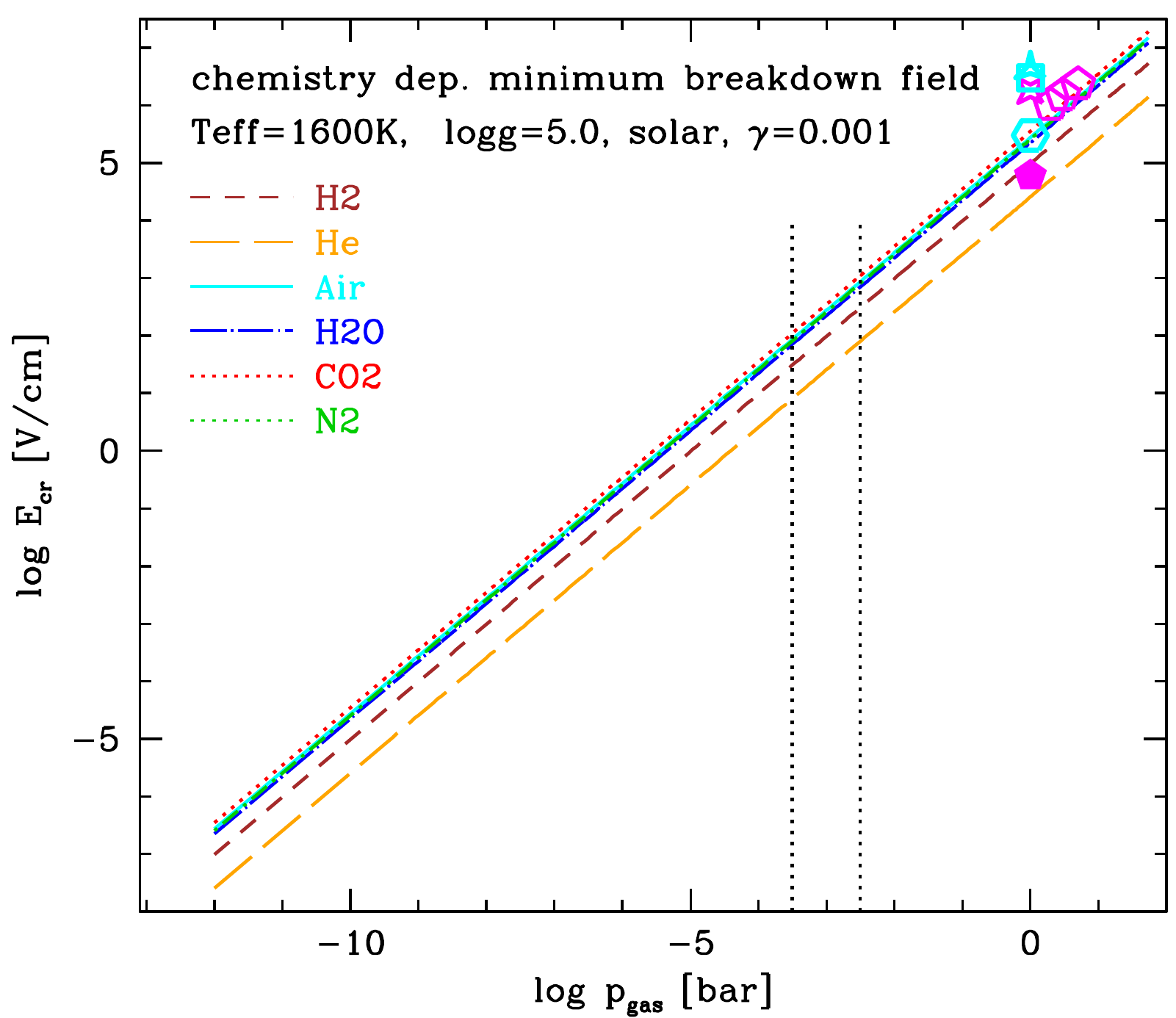}}
\caption{\small The minimum critical breakdown electric field
corresponding to the minimum of the Paschen curve for individual
ionised gases (H$_2$, He, air, H$_2$O, CO$_2$; Table 4.1 in Raizer
1991).  Pressures higher than indicated by the vertical dotted line
correspond to $d<6$cm, and pressures lower than indicated by the
vertical dotted line correspond to $d>6$cm according to
Fig.~\ref{fig:capdist_dg}. (This set of dotted double lines results
from the intersection interval of the dotted line in
Fig.~\ref{fig:capdist_dg}.)  The symbols indicate the breakdown field for
Earth (cyan) and Jupiter (magenta) as given in Yair et al. 1997 (Table
1; open symbols: measured, solid: with drops and ice, stars:
theroretical estimate).}
\label{fig:Etmin}
\end{figure}

\begin{figure}
\hspace*{1.3cm}\resizebox{10.5cm}{!}{\includegraphics{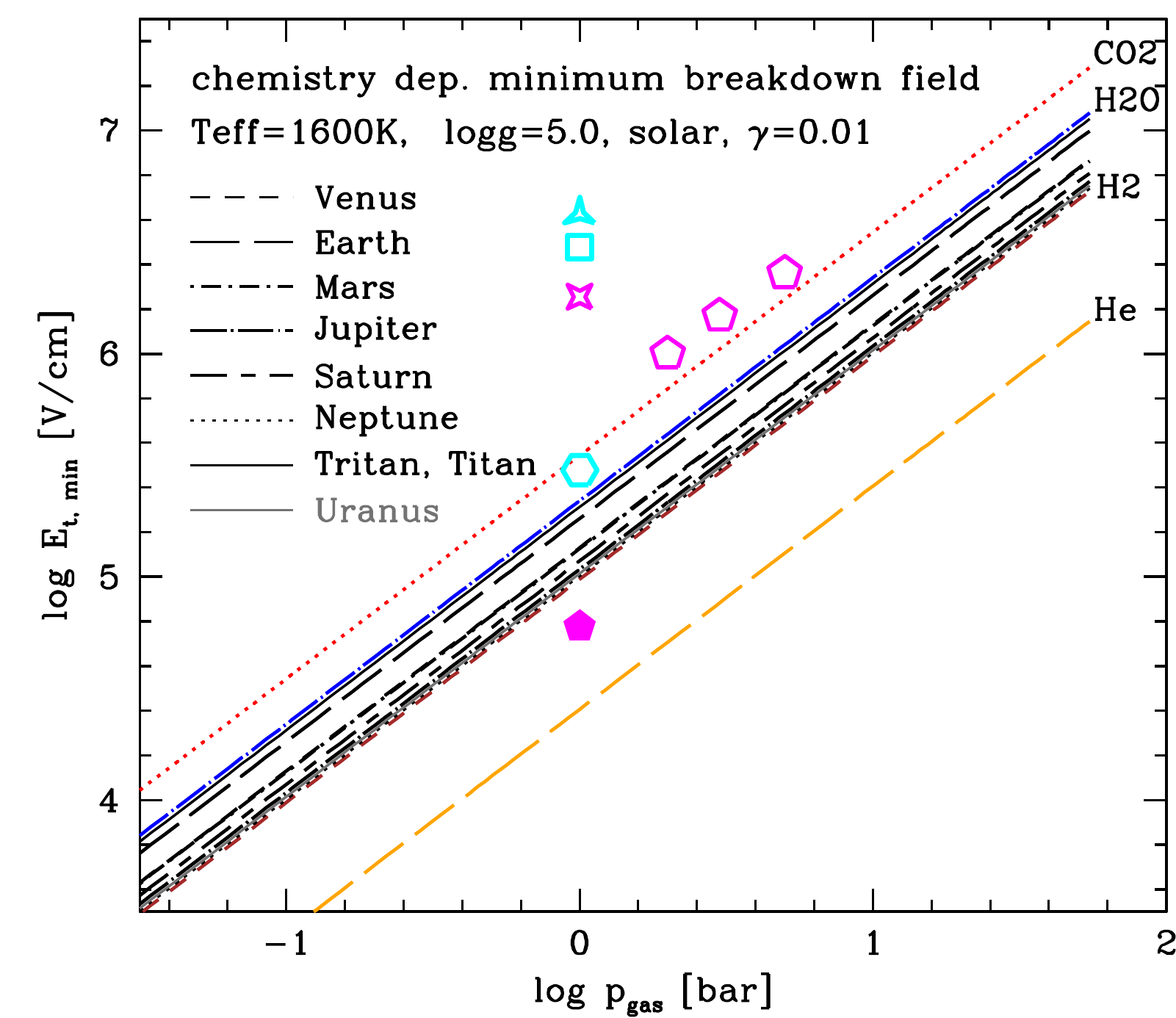}}
\caption{\small The minimum critical breakdown electric field for a
Brown Dwarf atmosphere with T$_{\rm eff}=1600$K and log(g)=5.0
assuming that the field break down happens in different gas
compositions. This is a zoom into Fig.~\ref{fig:Etmin} for CO$_2$ (red
dotted), H$_2$O (blue long-dashed), H$_2$ (brown dashed) and He
(orange long dashed) with the curves assuming the solar system planetary
atmosphere compositions over-plotted in black.  The curve for Uranus
and Neptune lie almost on top of each other.}
\label{fig:Etmin_zoom}
\end{figure}

\begin{figure}
\hspace*{1.3cm}\resizebox{12.5cm}{!}{\includegraphics{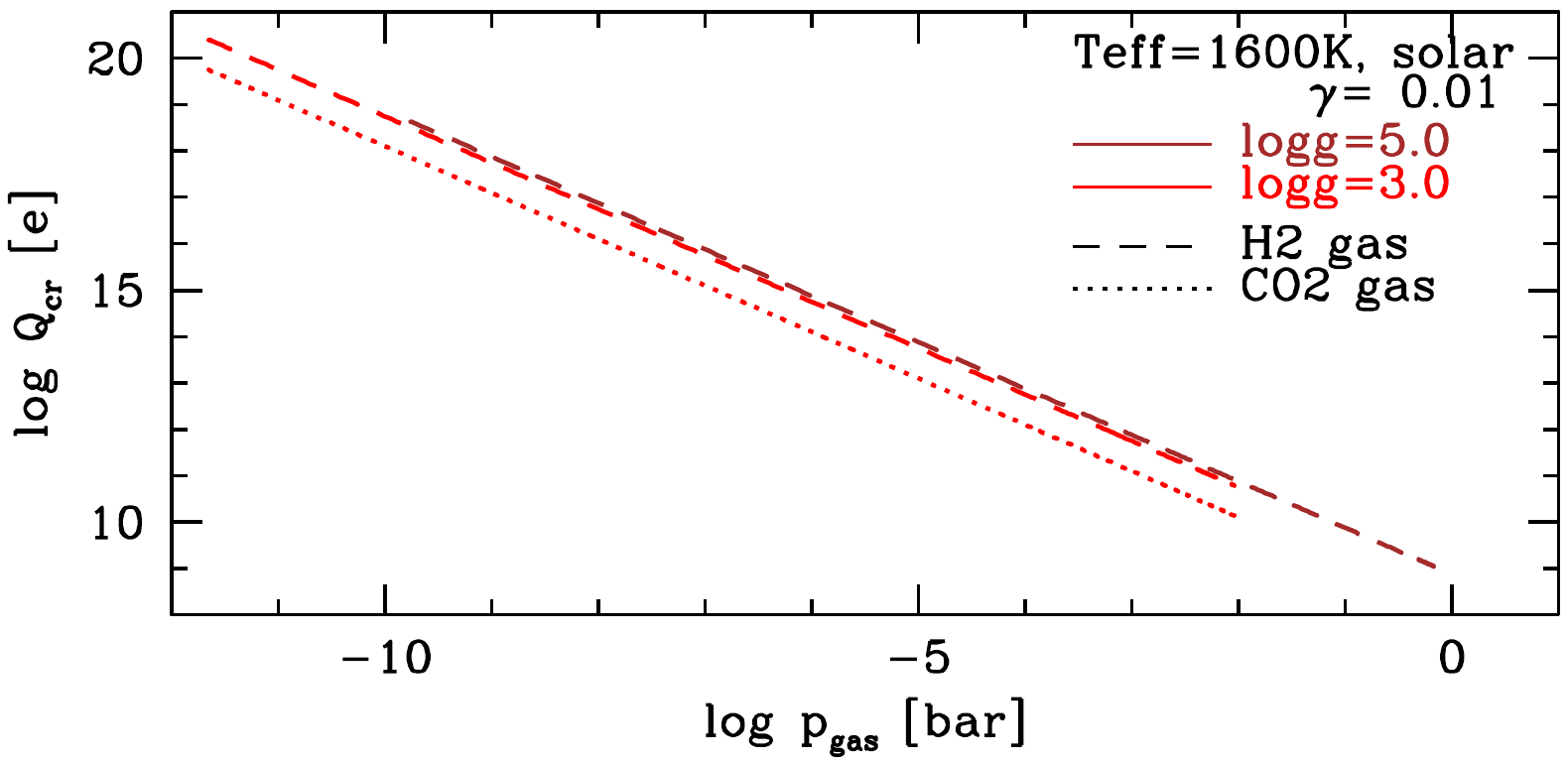}}\\
\hspace*{1.3cm}\resizebox{12.5cm}{!}{\includegraphics{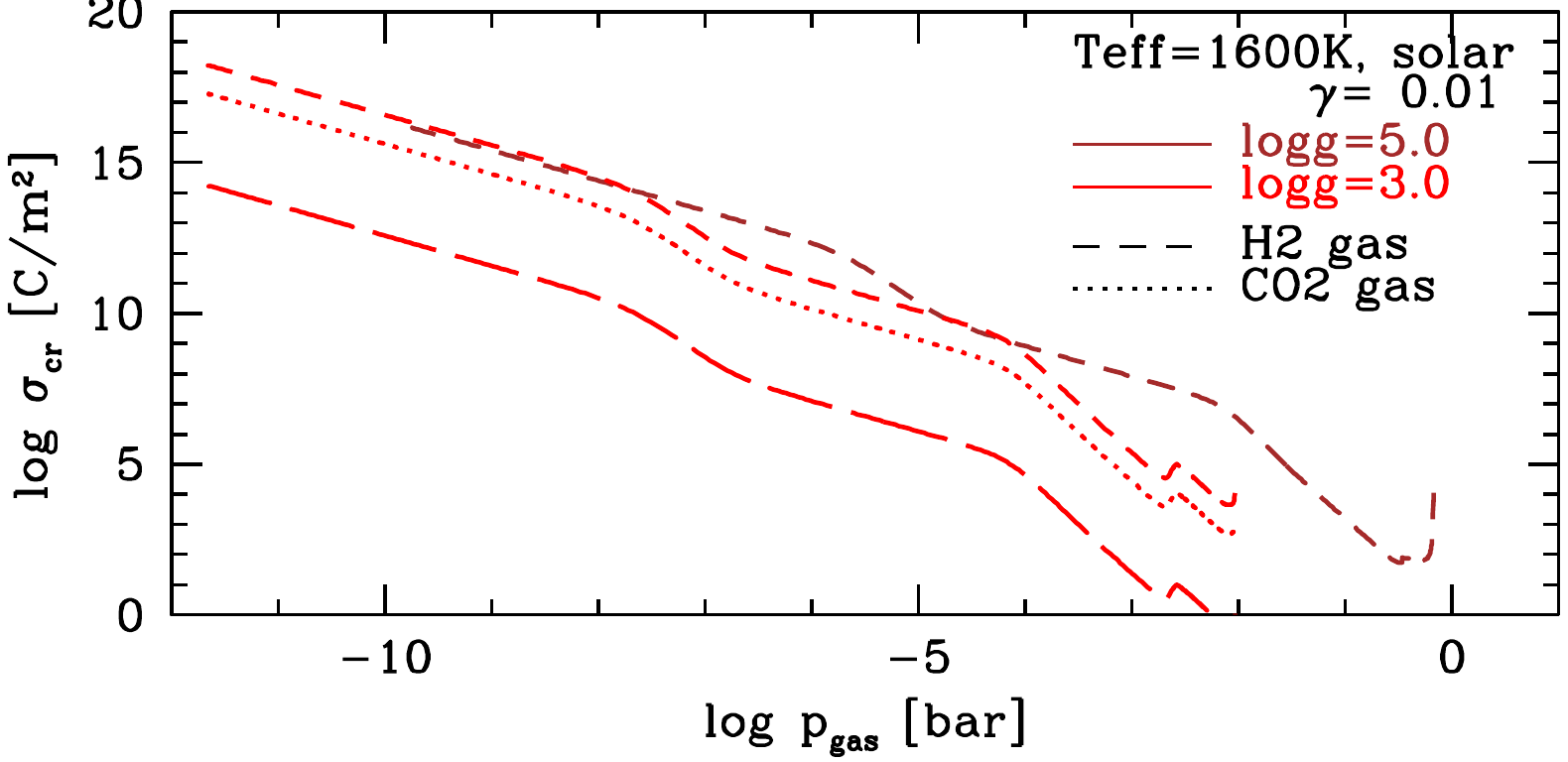}}
\caption{\small {\bf Top:} The total number of charges needed to
overcome the breakdown field (Eq.~\ref{eq:Qcrit}) for a H$_2$
dominated gas in a mineral cloud in a brown dwarf atmosphere (brown)
and a giant gas planet (red). These charges are assumed to exist on
two charge-carrying surfaces separated by the minimum distance for
discharge. The case of a CO$_2$-dominated atmosphere is over-plotted
for the giant-gas planet (dotted line). {\bf Bottom:} The critical surface
charge density, $\sigma_{\rm cr}=Q_{\rm cr}/(4\pi r^2)$ critical for
an electric field breakdown. $\sigma_{\rm cr}$ was calculated for
$r=<\!\!a(T,\rho_{\rm gas})\!\!>$ assuming that all grains are of the size of the local mean
grain size, except for the long-dashed line which shows $\sigma_{\rm
cr}$ for $100\times\!\!<\!\!a\!\!>$. Note that 1 C/m$^2 = 6.24\cdot 10^{14}$
e/cm$^2$. }
\label{fig:Qcrit}
\end{figure}

\begin{figure}
\hspace*{1.2cm}\resizebox{11.1cm}{!}{\includegraphics{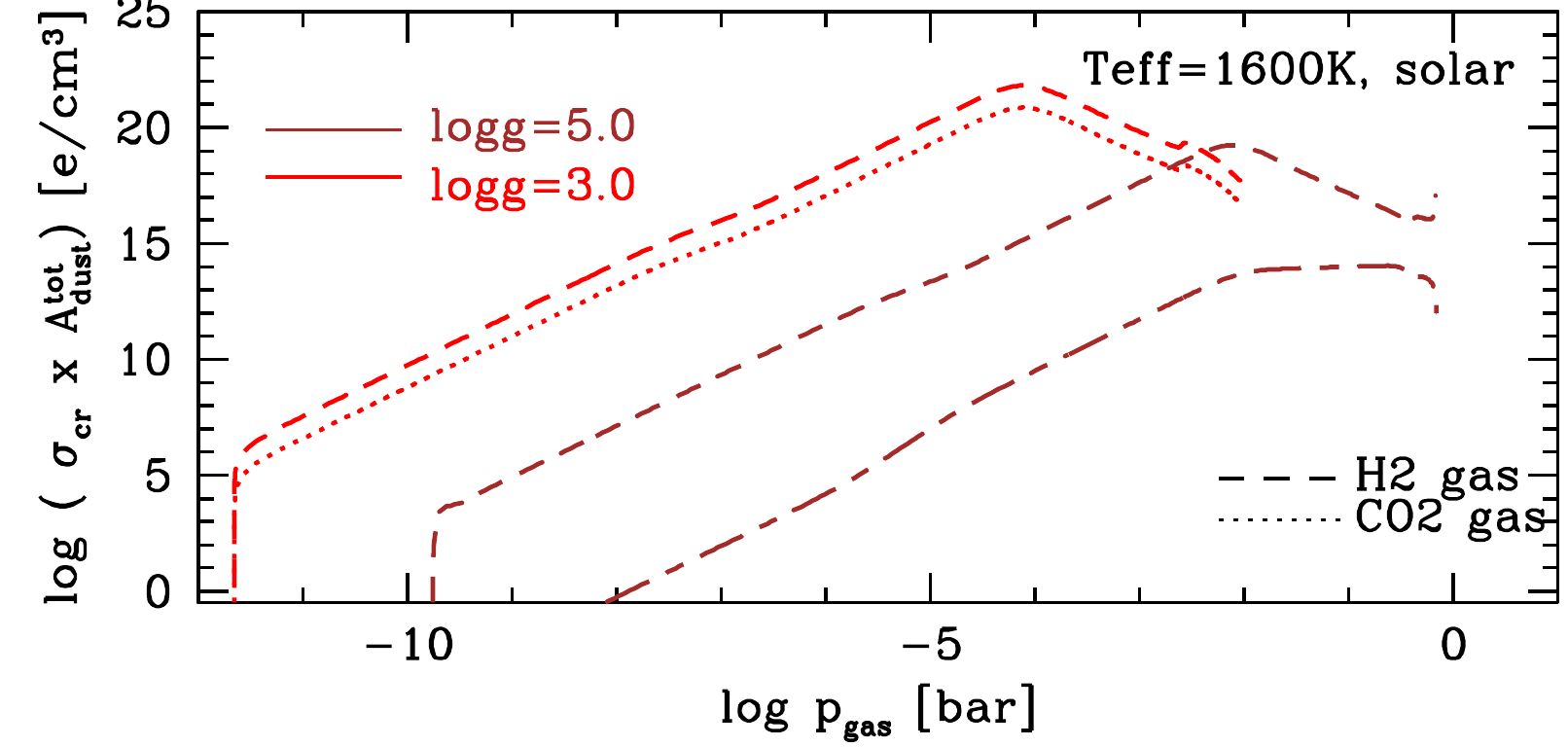}}\\
\hspace*{1.2cm}\resizebox{11.1cm}{!}{\includegraphics{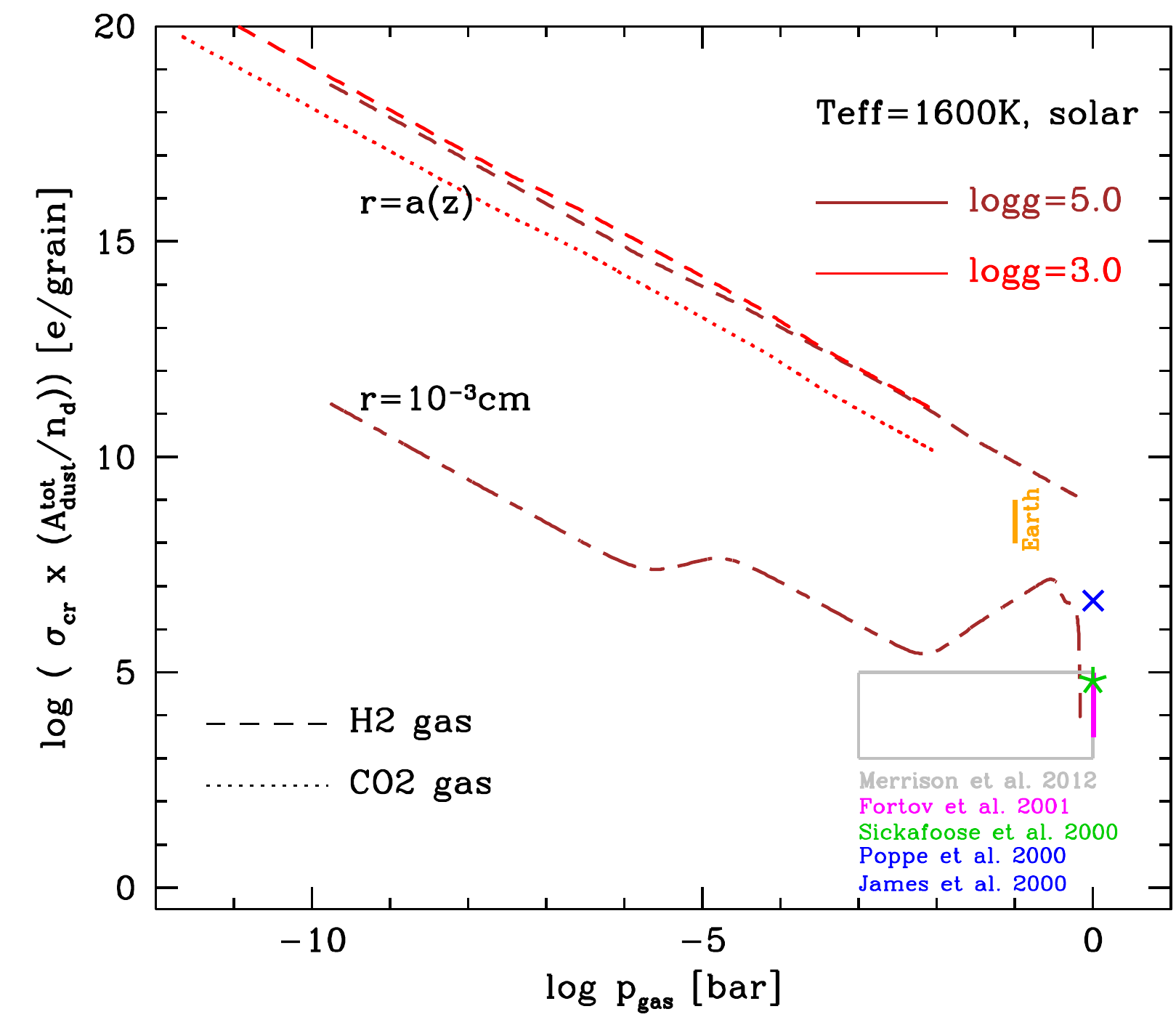}}
\caption{\small Critical number of charges needed for  electric
  breakdown in mineral clouds inside atmospheres for objects with
  T$_{\rm eff}$=1600K and log(g)=3.0 (red) and log(g)=5.0
  (brown). {\bf Upper panel:} The charges on the total dust surface
  $A^{\rm tot}_{\rm dust}$ per cm$^3$ of atmospheric gas as
  $\sigma_{\rm cr} \times A^{\rm tot}_{\rm dust}$ [e/cm$^3$]. {\bf
    Lower panel:} The number of charges per dust grain $\sigma_{\rm
    cr} \times (A^{\rm tot}_{\rm dust}/n_{\rm d})$ [e/grain].  The symbols represent the following data sources: grey rectangle -- Merrison et al. (2012), magenta vertical bar -- Fortov et al. (2001), green asterisk -- Sickafoose et al. (2000), blue cross -- Poppe et al. (2000) and James et al. (2000), orange vertical bar -- values for Earth hale from Lamb \& Verlinde (2011).}
\label{fig:QcritxA}
\end{figure}

\begin{figure}
\hspace*{1.3cm}\resizebox{10.5cm}{!}{\includegraphics{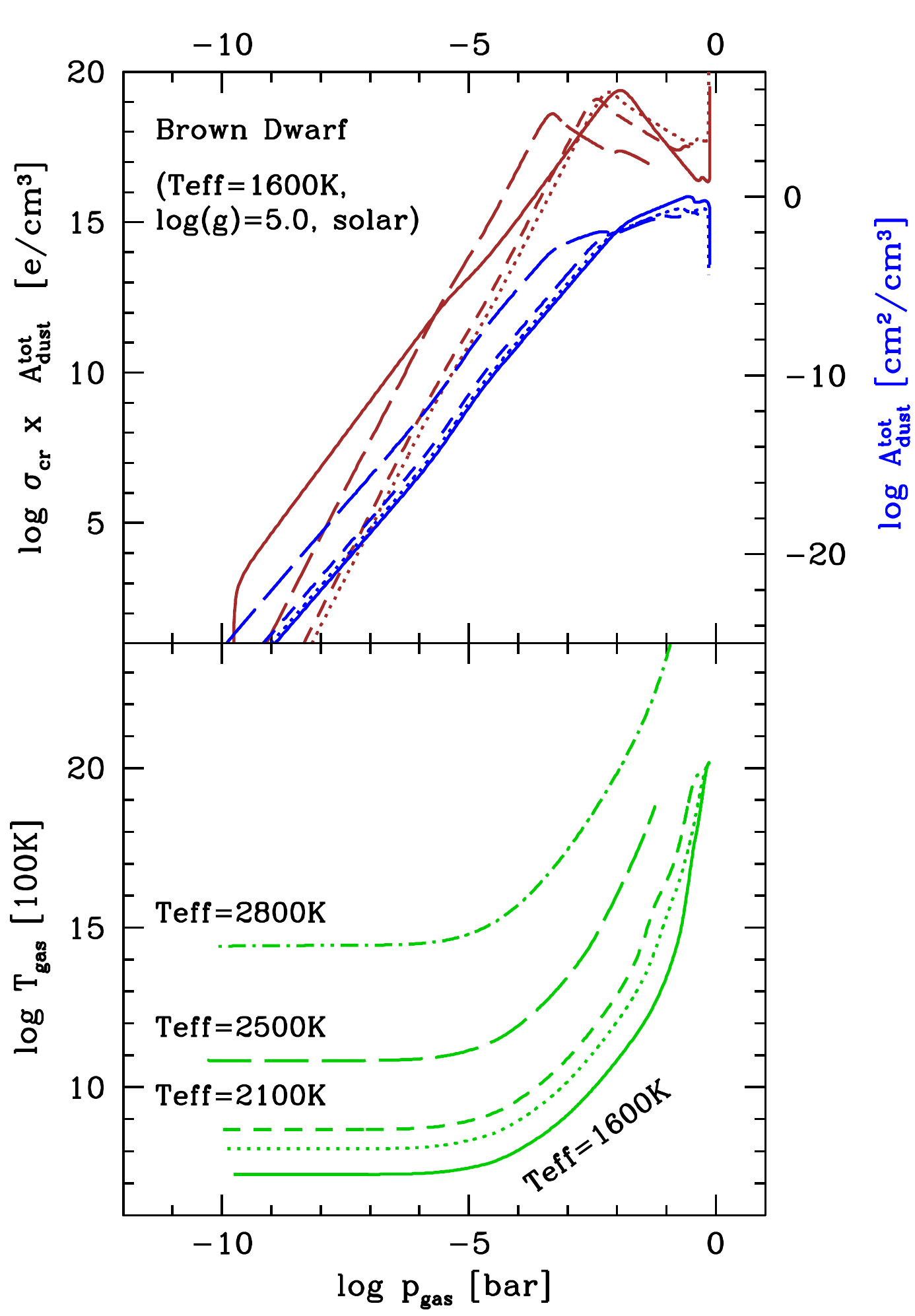}}
\caption{\small Critical charge number density, $\sigma_{\rm cr}$ [e/cm$^3$] per total  dust surface needed to overcome the local breakdown
field depending on the effective temperature of the object (top
panel). The results are plotted for 5 solar-metalicity Brown Dwarf
model atmospheres (log(g)=5.0) with T$_{\rm eff}=1600$K (solid), 1900K
(dotted), 2100K (short dashed), 2500K (long dashed), 2800K
(dash-dot). The  lower panel (local gas temperature) demonstrates
why the critical surface charge density $\sigma_{\rm cr}(p)$ changes for different effective temperatures. }
\label{fig:Qcrit_Teff}
\end{figure}

\begin{figure}
\hspace*{1.3cm}\resizebox{10.5cm}{!}{\includegraphics{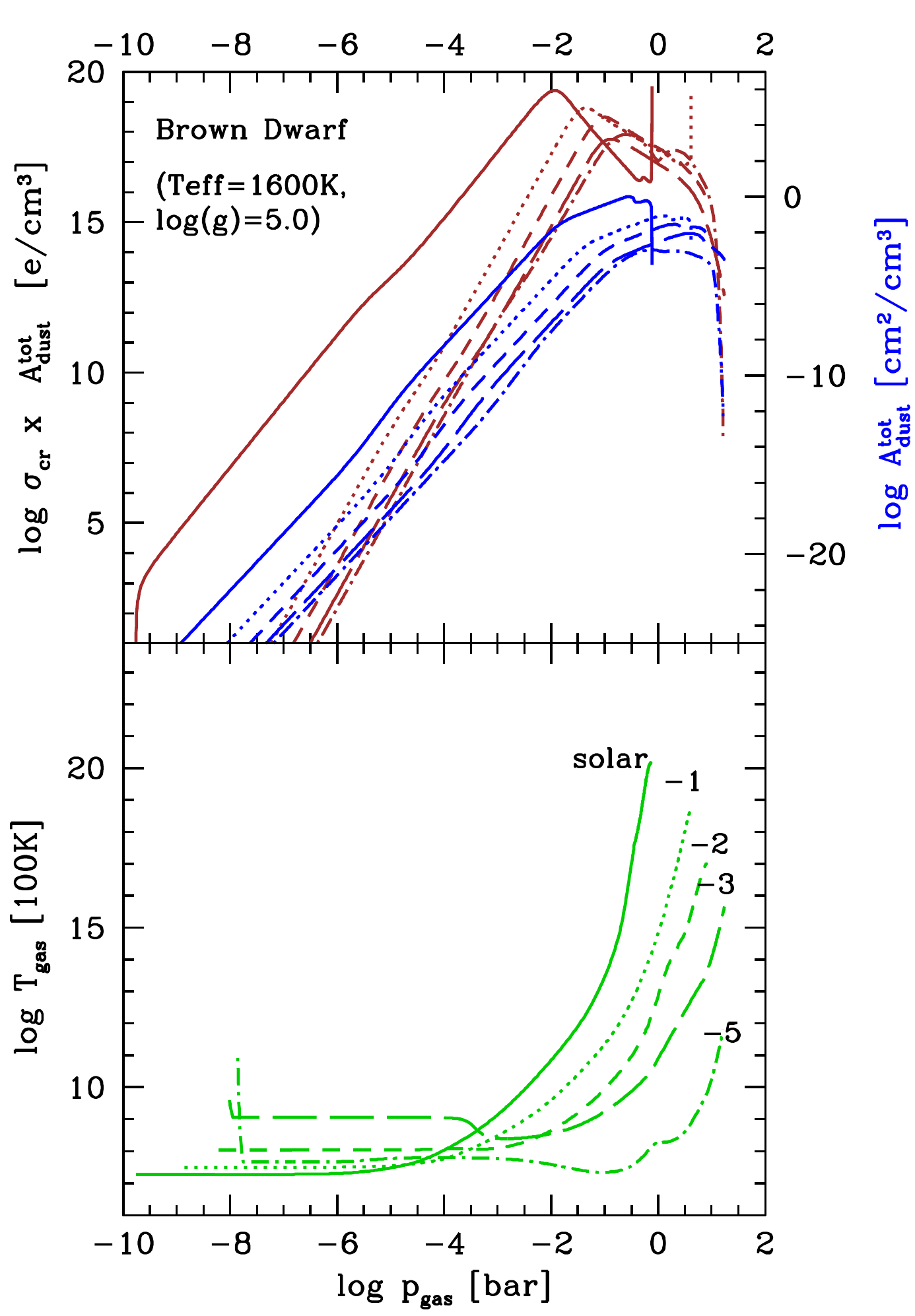}}
\caption{\small Critical charge number density, $\sigma_{\rm cr}$ [e/cm$^3$] per total dust surface, needed to overcome the local breakdown
field depending on the metalliciy of the object (top panel). The results
are plotted for 5 Brown Dwarf model atmospheres (log(g)=5.0) with
different metallicities [M/H]=0 (solar, solid), -1 (dotted), -1 (short
dashed), -3 (long dashed), -5 (dash-dot). The lower panel (local
gas temperature) demonstrate why  the critical surface charge density  $\sigma_{\rm cr}(p)$ changes for different
[M/H]. }
\label{fig:Qcrit_metal}
\end{figure}

\begin{figure}
\hspace*{1.3cm}\resizebox{12.5cm}{!}{\includegraphics{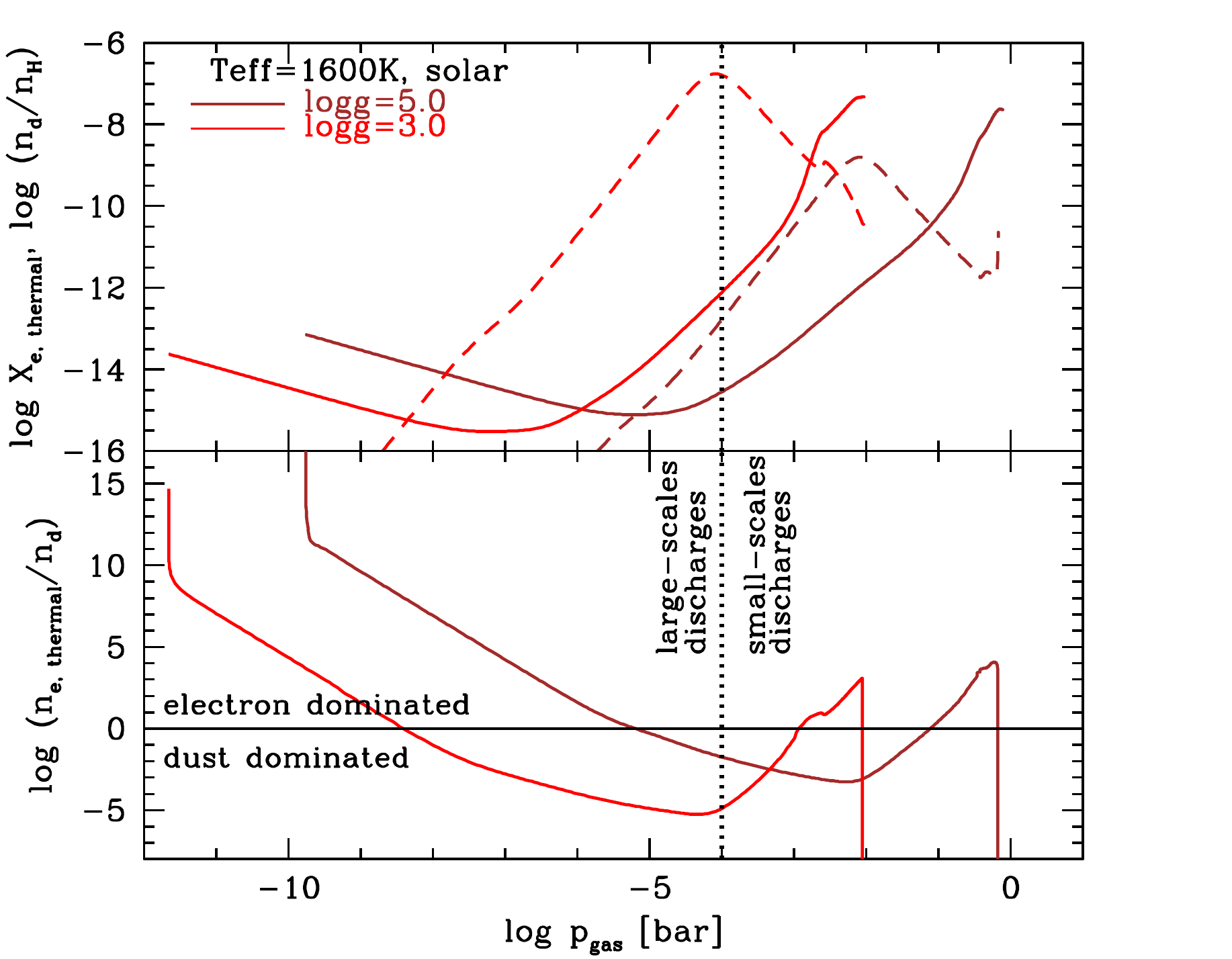}}
\caption{\small The efficiency of discharges in planetary atmospheres
with mineral clouds. {\bf Top:} Degree of thermal gas ionisation,
$\chi_{\rm e, themal}=n_{\rm e}/n_{\rm H}$ (solid lines), and the dust
number density per hydrogen, $n_{\rm d}/n_{\rm H}$ (dashed lines), for
a Brown Dwarf and a giant gas planet atmosphere for the same
Drift-Phoenix models as in Fig.~\ref{fig:Qcrit}. {\bf Bottom:} The
ratio of the number of thermal electrons, $n_{\rm e, thermal}$, to the
number of cloud particles, $n_{\rm d}$, provides a measure of the
efficiency with which a streamer may be initiated in a mineral
cloud. Two different regimes appear: above the black line - electron
dominated, hence more than one electron is available per grain pair,
below the black line - dust dominated, hence less than one electron is
available per grain pair. The dotted vertical lines distinguished the
cloud regimes of potential large-scale cloud discharges and
small-scale, inter-grain discharges similar to
Figs.~\ref{fig:capdist}\,-\,\ref{fig:Etmin}}
\label{fig:fracide}
\end{figure}

\begin{figure}
\hspace*{1.3cm}\resizebox{12.5cm}{!}{\includegraphics{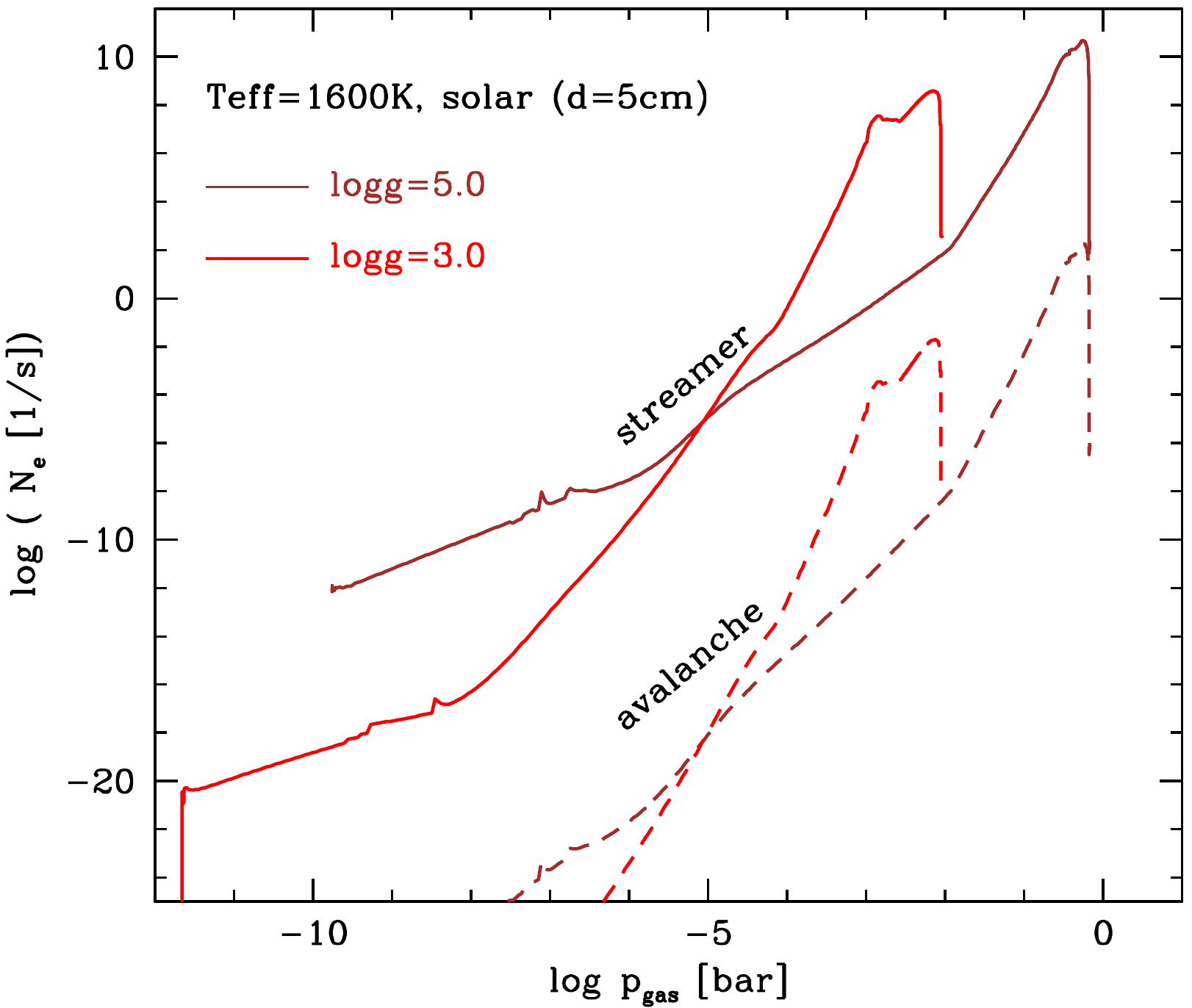}}
\caption{\small Gas-phase enrichment by dust-dust collision
triggered discharges.  Electron avalanche (dashed lines; $d=5$cm
used) yields considerably less free charges than streamer events
(solid lines).  The same Brown Dwarf (brown) and a giant gas planet
(red) atmosphere models are used as in Fig.~\ref{fig:fracide}.}
\label{fig:Nestream}
\end{figure}

\clearpage

\end{document}